\documentclass[english]{article}
\usepackage{lmodern}

\usepackage[T1]{fontenc}
\usepackage[latin9]{inputenc}
\usepackage{booktabs}
\usepackage{mathrsfs}
\usepackage{amsmath}
\usepackage{graphicx}

\makeatletter

\providecommand{\tabularnewline}{\\}

\newcommand{\lyxaddress}[1]{
\par {\raggedright #1
\vspace{1.4em}
\noindent\par}
}

\usepackage{mathrsfs}   
\usepackage{slashed}     
\usepackage{bbold}  
\usepackage{url}
\usepackage{graphicx}
\usepackage[colorlinks=true,linkcolor=redLinks,citecolor=greenLinks,urlcolor=redLinks, pdfborder={0 0 1}]{hyperref}
\usepackage{xcolor}
\usepackage{framed}
\usepackage[numbers,sort&compress]{natbib}
\usepackage{amsmath}
\usepackage{pifont}

\allowdisplaybreaks

\colorlet{shadecolor}{gray!15}

\definecolor{greenLinks}{rgb}{0, 0.6, 0} 
\definecolor{blueLinks}{rgb}{0, 0, 0.6}
\definecolor{redLinks}{rgb}{0.6, 0, 0}
\definecolor{tempText}{rgb}{0.55, 0.10,0.67}
\definecolor{eprintLinks}{rgb}{0.4, 0.4, 0.4}
\definecolor{journalLinks}{rgb}{0.6, 0, 0}

\newcommand{\MYhref}[3][redLinks]{\href{#2}{\color{#1}{#3}}}%

\def\gsim{\raise0.3ex\hbox{$\;>$\kern-0.75em\raise-1.1ex\hbox{$\sim\;$}}}
\def\lsim{\raise0.3ex\hbox{$\;<$\kern-0.75em\raise-1.1ex\hbox{$\sim\;$}}}

\usepackage{multirow}
\let\orig@Hy@EveryPageAnchor\Hy@EveryPageAnchor
\def\Hy@EveryPageAnchor{%
    \begingroup
    \hypersetup{pdfview=Fit}%
    \orig@Hy@EveryPageAnchor
    \endgroup
}

\let\oldFootnote\footnote
\newcommand\nextToken\relax

\renewcommand\footnote[1]{%
    \oldFootnote{#1}\futurelet\nextToken\isFootnote}

\newcommand\isFootnote{%
    \ifx\footnote\nextToken\textsuperscript{,}\fi}

\definecolor{myPurple}{RGB}{128,0,182}

\newcommand{\cmark}{\ding{51}}%
\newcommand{\xmark}{\ding{55}}%

\date{}

\makeatother

\usepackage{babel}
\begin{document}

\title{{\Large{}\vspace{-1.0cm}} \hfill {\normalsize{}IFIC/16-34} \\*[10mm] 
Lepton number violation in 331 models}


\author{{\Large{}Renato M. Fonseca}\thanks{E-mail: renato.fonseca@ific.uv.es} 
\ and {\Large{}Martin Hirsch}\thanks{E-mail: mahirsch@ific.uv.es} 
\date{}}

\maketitle

\lyxaddress{\begin{center}
{\Large{}\vspace{-0.5cm}}\href{http://www.astroparticles.es/}{AHEP Group},
\href{http://webific.ific.uv.es/web/en}{Instituto de Física Corpuscular},
\href{http://www.csic.es}{C.S.I.C.}/\href{http://www.uv.es/}{Universitat de València}\\
Parc Científic de Paterna.  Calle Catedrático José Beltrán, 2 E-46980
Paterna (Valencia) -- Spain
\par\end{center}}

\begin{center}
\today
\par\end{center}
\begin{abstract}
Different models based on the extended $SU(3)_{C}\times
SU(3)_{L}\times U(1)_{X}$ (331) gauge group have been proposed over
the past four decades. Yet, despite being an active research topic,
the status of lepton number in 331 models has not been fully
addressed in the literature, and furthermore many of the original proposals
can not explain the observed neutrino masses.  In this paper we
review the basic features of various 331 models, focusing on potential sources of
lepton number violation. We then describe different modifications
which can be made to the original models in order to accommodate 
neutrino (and charged lepton) masses.

\end{abstract}

\noindent 
{\vfill
}

\pagebreak{}

\section{Introduction}

It is conceivable that the Standard Model gauge symmetry
$SU(3)_{C}\times SU(2)_{L}\times U(1)_{Y}$ (321) is just a remnant of
a larger one. Indeed, such scenarios are attractive as they are able
to unify the three gauge couplings, provided that the extended gauge
group is simple
\cite{Georgi1974,Georgi1975,Fritzsch1975,Guersey1976}. However, one
should not exclude the possibility that the enlarged group is a
product of simple factors. This could happen as an intermediate step
towards a grand unified group.  A famous example is the left-right
symmetric group $SU(3)_{C}\times SU(2)_{L}\times SU(2)_{R}\times
U(1)_{B-L}$ \cite{Pati:1974yy,Mohapatra:1974gc,Mohapatra:1980yp},
which fits neatly into $SO(10)$.  Another possibility is
$SU(3)_{C}\times SU(3)_{L}$, yet with such models one cannot get the
correct fermion masses \cite{Fonseca:2015aoa}. On the other hand, it
was realized long ago \cite{Singer1980,Pisano1992,Frampton1992} that
with an extra $U(1)_{X}$ it is possible to construct viable models.

These $SU(3)_{C}\times SU(3)_{L}\times U(1)_{X}$ (331) models have
received considerable attention in connection with various topics:
neutrino mass generation
\cite{Kitabayashi2001,Tully2001,Boucenna2014,Boucenna2015,Pires2014,Okada2016,Okamoto1999,Montero:2001tq,Montero:2001ts,Montero:2004uy,Okada:2015bxa,Dong:2006mt,Kitabayashi:2001dg,Long1996,Foot1994,Chang:2006aa,Gutierrez:2004sba,TavaresVelasco:2004be,Long:1999ij,Mizukoshi:2010ky},
flavour symmetries
\cite{Hernandez:2015tna,Hernandez:2015cra,Vien:2014gza,Hernandez:2014lpa,Hernandez:2014vta,Vien:2013zra,Yin:2007rv,Dong:2011vb,Dong:2010zu,Dong:2010gk,Dias:2003iq,Hernandez:2013hea},
quark flavour 
observables \cite{Buras2008,Buras2014,Gonzalez-Sprinberg2005,Agrawal:1995vp,Buras:2012dp,Promberger:2007py,Rodriguez:2004mw,Buras:2014yna} or 
the recent LHC diphoton excess
\cite{Boucenna2015a,Dong2015,Cao2016,Mantilla:2016sew,Hernandez:2015ywg,Martinez:2016ztt},
among others. Underpinning this interest is the fact that the 331 to
321 symmetry breaking energy scale can be of the TeV order, hence it
could possibly be explored at the LHC; see for example 
\cite{Coutinho:2013lta,Montalvo:2013tra,Richard:2013xfa,Pelaggi:2015knk,Dong:2015dxw,Nepomuceno:2016jyr,Coriano:2016itf,Okada:2016whh}. 

However, despite the large list of papers on 331 models, the issue of
lepton number violation (LNV) has not been fully addressed in the
literature and, in fact, many misleading statements on the subject can
be found in papers on 331 models.  
It turns out that models based on this extended
symmetry can be quite different from one another since the way the
321 group is embedded in the 331 group is not unique. In particular (a)
the existence of neutrino masses, (b) the nature of neutrinos and (c)
the status of lepton number varies markedly among 331 models. As such,
with this work we intend to collect and summarize the relevant
information concerning lepton number and neutrino mass generation in
this class of models.

We have found that several of the originally proposed 331 models can
not explain correctly the observed neutrino masses (nor charged
lepton masses, in one case). Thus, it is necessary to extend these models and 
we present several possible modifications that can bring these 
models in agreement with experimental data, some of which have already been considered before \cite{Foot1993,Liu:1993gy,Ng:1992st,Duong:1993zn,Montero:2001ji,Boucenna2014}. We focus here (mostly) 
on neutrino masses and mixings and leave aside other 
LNV processes, which we mention only briefly when it is relevant. 

The rest of this paper is organized as follows.  Section \ref{sec:2}
describes the basic features of six different 331 models. Four of
these fall into a particular subclass since they have a common structure
(they all follow what we call the SVS framework, after its prototype model
\cite{Singer1980}). To cover the
full variety of 331 models, we then discuss two more models,
which do not follow the SVS scheme, and clarify LNV related issues in
them as well.  None of the basic models in the SVS
class generates lepton masses and mixings in a fully satisfactory way,
hence modifications are required.  A list of simple improvements is
discussed in section \ref{sec:3}. For each of the possibilities in our
list we give a brief description on how the modified versions of the
original models can be brought into agreement with experimental
neutrino (and charged lepton) data.  Finally, in section \ref{sec:4}
we summarize the most important points in this manuscript. An appendix
at the end of the text provides supplementary information.

\section{\label{sec:2}The $SU(3)_{L}\times U(1)_{X}$ group and basic 331
	models}

One can build different 331 models, not just by changing the field
content, but also by varying the way in which the SM electroweak gauge
group is embedded in $SU(3)_{L}\times U(1)_{X}$. This can be encoded
in a continuous parameter $\beta$ which controls the relation between
the hypercharge $Y$, $X$, and the $T_{8}$ generator of $SU(3)_{L}$:
\begin{align}
	Y & =\beta T_{8}+X\,.
\end{align}
From here one can derive that $SU(3)_{L}\times U(1)_{X}$
representations break as follows into $SU(2)_{L}\times U(1)_{Y}$
representations (more details can be found in the appendix):\footnote{Hats
	are added to $SU(2)_{L}$ representations to avoid
	confusion between 331 and 321 representations.}
\begin{align}
	\left(\mathbf{3},x\right) & \rightarrow\left(\widehat{\mathbf{2}},x+\frac{1}{2\sqrt{3}}\beta\right)+\left(\widehat{\mathbf{1}},x-\frac{1}{\sqrt{3}}\beta\right)\,,\label{eq:triplet_decomposition}\\
	\left(\overline{\mathbf{6}},x\right) & \rightarrow\left(\widehat{\mathbf{3}},x-\frac{1}{\sqrt{3}}\beta\right)+\left(\widehat{\mathbf{2}},x+\frac{1}{2\sqrt{3}}\beta\right)+\left(\widehat{\mathbf{1}},x+\frac{2}{\sqrt{3}}\beta\right)\,,\\
	\left(\mathbf{8},0\right) & \rightarrow\left(\widehat{\mathbf{3}},0\right)+\left(\widehat{\mathbf{1}},0\right)+\left(\widehat{\mathbf{2}},-\frac{\sqrt{3}}{2}\beta\right)+\left(\widehat{\mathbf{2}},\frac{\sqrt{3}}{2}\beta\right)\,.\label{eq:octet_decomposition}
\end{align}
Together with the requirement $\left|\beta\right|<\tan^{-1}\theta_{W}\approx1.8$ (obtainable from equation \eqref{eq:gY} and the fact that $g^2_X$ must be positive),
these equations show that there are only four values of $\beta$ for
which it is possible to avoid colorless, fractionally charged
fermions. Bearing this constraint in mind, we can then describe
six different 331 models:
\begin{itemize}
	\item In the first four models, the three lepton families are in equal
	representations, but the quarks are not. The structure of these
	models is similar, with the main difference between them being the
	value of $\beta$: $-1/\sqrt{3}$ in the Singer-Valle-Schechter (SVS)
	model \cite{Singer1980}, $-\sqrt{3}$ in the Pisano-Pleitez-Frampton (PPF) model
	\cite{Pisano1992,Frampton1992},
	$1/\sqrt{3}$ in the Pleitez-Özer model \cite{Pleitez:1994pu,Ozer:1995xi}, and $\sqrt{3}$ in what we call the 
	model X. They all share a common structure, which we call the SVS framework below.
	\item The flipped model \cite{Fonseca:2016tbn}, where quark families are all in the same
	representations, but leptons are not.
	\item The $E_{6}$ model \cite{Sanchez:2001ua}, where complete family replication is true for
	both the lepton and quark sectors.
\end{itemize}

\setlength\tabcolsep{4pt}

\subsection{The Singer-Valle-Schechter (SVS) model}

The first 331 model with three generations of quarks and leptons
was proposed in \cite{Singer1980}, using $\beta=-1/\sqrt{3}$. As stated previously, this 
model can be considered the prototype model for what might be called 
the SVS framework. All four models in this class have in 
common the following features:
\begin{itemize}
	\item The SM lepton doublets are placed in three triplets of $SU(3)_{L}$.\footnote{
		The original model in \cite{Singer1980} also contained right-handed neutrinos
		in $SU(3)_{L}$ singlets, which were latter removed \cite{Valle:1983dk}. Here, we call this variation of the original proposal "the SVS model".}
	\item Two families of left-handed quarks are placed inside anti-triplets
	of $SU(3)_{L}$ while the third one is placed in a triplet.
	\item Extra $SU(3)_{L}$ fermion singlets are necessary in order to
	include some of the SM $SU(2)_{L}$ singlets, and also to provide the
	necessary vector partners to some extra fermions contained in the
	triplets of $SU(3)_{L}$.
	\item Three scalar triplets of $SU(3)_{L}$ are used to generate the
	necessary Yukawa interactions with fermions.
\end{itemize}
These conditions guarantee that models in this class recover correctly
the SM fermion content in the limit where 331 is first broken to 321,
and they also have a sufficiently large scalar sector to achieve both 331
symmetry breaking and a realistic quark spectrum.

In the specific case of the SVS model where $\beta=-1/\sqrt{3}$,
right-handed neutrinos, here denoted $N^c$, are included in the same
extended gauge multiplet $\psi_{\ell}$ as the SM left-handed leptons. The full
field content of the original SVS model is shown in table
\ref{tab:SVS-reps}. In addition to the SM fermions, extra vector-like 
quarks appear, which are a common feature of all 331 models. 

\begin{table}[tbph]
	\begin{centering}
		\begin{tabular}{cccccc}
			\toprule 
			Field & 331 representation & $G_{SM}$ decomposition & \# flavours & Components & Lepton number\tabularnewline
			\midrule
			$\psi_{\ell,\alpha}$ & $\left(\mathbf{1},\mathbf{3},-\frac{1}{3}\right)$ & $\left(\mathbf{1},\widehat{\mathbf{2}},-\frac{1}{2}\right)+\left(\mathbf{1},\widehat{\mathbf{1}},0\right)$ & 3 & $\left(\left(\nu_{\alpha},\ell_{\alpha}\right),N_{\alpha}^{c}\right)^{T}$ & $\left(1,1,-1\right)^{T}$\tabularnewline
			$\ell_{\alpha}^{c}$ & $\left(\mathbf{1},\mathbf{1},1\right)$ & $\left(\mathbf{1},\widehat{\mathbf{1}},1\right)$ & 3 & $\ell_{\alpha}^{c}$ & $-1$\tabularnewline
			$Q_{\alpha=1,2}$ & $\left(\mathbf{3},\overline{\mathbf{3}},0\right)$ & $\left(\mathbf{3},\widehat{\mathbf{2}},\frac{1}{6}\right)+\left(\mathbf{3},\widehat{\mathbf{1}},-\frac{1}{3}\right)$ & 2 & $\left(\left(d_{\alpha},-u_{\alpha}\right),D_{\alpha}\right)^{T}$ & $\left(0,0,2\right)^{T}$\tabularnewline
			$Q_{3}$ & $\left(\mathbf{3},\mathbf{3},\frac{1}{3}\right)$ & $\left(\mathbf{3},\widehat{\mathbf{2}},\frac{1}{6}\right)+\left(\mathbf{3},\widehat{\mathbf{1}},\frac{2}{3}\right)$ & 1 & $\left(\left(t,b\right),U\right)^{T}$ & $\left(0,0,-2\right)^{T}$\tabularnewline
			$u_{\alpha}^{c}$ & $\left(\overline{\mathbf{3}},\mathbf{1},-\frac{2}{3}\right)$ & $\left(\overline{\mathbf{3}},\widehat{\mathbf{1}},-\frac{2}{3}\right)$ & 4 & $u_{\alpha}^{c}$ & 0\tabularnewline
			$d_{\alpha}^{c}$ & $\left(\overline{\mathbf{3}},\mathbf{1},\frac{1}{3}\right)$ & $\left(\overline{\mathbf{3}},\widehat{\mathbf{1}},\frac{1}{3}\right)$ & 5 & $d_{\alpha}^{c}$ & 0\tabularnewline
			$\phi_{1}$ & $\left(\mathbf{1},\mathbf{3},\frac{2}{3}\right)$ & $\left(\mathbf{1},\widehat{\mathbf{2}},\frac{1}{2}\right)+\left(\mathbf{1},\widehat{\mathbf{1}},1\right)$ & 1 & $\left(\left(\phi_{1}^{+},\phi_{1}^{0}\right),\widetilde{\phi}_{1}^{+}\right)^{T}$ & $\left(0,0,-2\right)^{T}$\tabularnewline
			$\phi_{i=2,3}$ & $\left(\mathbf{1},\mathbf{3},-\frac{1}{3}\right)$ & $\left(\mathbf{1},\widehat{\mathbf{2}},-\frac{1}{2}\right)+\left(\mathbf{1},\widehat{\mathbf{1}},0\right)$ & 2 & $\left(\left(\phi_{i}^{0},\phi_{i}^{-}\right),\widetilde{\phi}_{i}^{0}\right)^{T}$ & $\left(0,0,-2\right)^{T}$\tabularnewline
			\bottomrule
		\end{tabular}
		\par\end{centering}
	
	\protect\caption{\label{tab:SVS-reps}Field content of the
		Singer-Valle-Schechter (SVS) model \cite{Singer1980}. The indices
		$\alpha$ and $i$ denote different flavours.}
\end{table}

To determine whether or not there is lepton number conservation in a
given model, one can simply attempt to build diagrams describing
processes where the number of leptons changes. Finding one such
diagram would prove conclusively that there is LNV. On the other hand,
if one is able to show that no such diagram exists, then lepton number is
preserved (perturbatively at least). The latter, however, can be 
quite cumbersome, when worked out with the language of Feynman 
diagrams.

In practice, thus, it is better to replace this pragmatic approach by
the following simpler one: show whether or not the total Lagrangian of
the model has a global $U(1)_{L}$ symmetry under which the SM
(anti)leptons have $+1$($-1$) charge, and (anti)quarks as well as the SM
gauge bosons have no charge.\footnote{We stress here that this $U(1)_{L}$
	does not need to commute
	with the remaining symmetries of the model (in particular, the
	$SU(3)_{L}\times U(1)_{X}$ gauge symmetry in 331 models).}
Lepton number is violated if and only if no such symmetry exists.  

If there is LNV, then usually there is no single coupling which is
responsible for it --- rather, it is the existence of several
interactions which gives rise to the phenomenon.  Nevertheless, in
practice only a few of the couplings in a given model are relevant for
LNV and in their absence, the Lagrangian gains a $U(1)_{L}$ symmetry
with the characteristics previously described.  However, this means
that one can have situations where the removal of either of two sets
of interactions --- $\left\{ I_{i}\right\} $, $\left\{ I'_{i}\right\}$ 
--- both lead to a lepton number conserving scenario, hence the
procedure of labeling LNV interactions is not unique, see below. 

Finally, one has to bear in mind that, even if the Lagrangian is
$U(1)_{L}$ preserving, it is still possible for lepton number to be
broken spontaneously by the vacuum expectation value (VEV)
of scalars which carry a non-zero $U(1)_{L}$ charge.

We now exemplify once the application of these well known (but often
neglected) comments, and derive the $U(1)_{L}$ charges in the
last column of table \ref{tab:SVS-reps}, which correspond to the SVS
model. For reasons which will become obvious later, we first put 
the coefficient of the term  $\phi_{1}\phi_{2}\phi_{3}$ to zero. 
Using the field notation in that table, we then may start from the
lepton Yukawa interactions $\psi_{\ell}\psi_{\ell}\phi_{1}$ and
$\psi_{\ell}\ell^{c}\phi_{1}^{*}$: from the first one it follows that
$L\left(\phi_{1}^{+}\right)=L\left(\phi_{1}^{0}\right) =
-1-L\left(N_{\alpha}^{c}\right)$ and
$L\left(\widetilde{\phi}_{1}^{+}\right)=-2$, while from the second
interaction we conclude that
$L\left(\phi_{1}^{+}\right)=L\left(\phi_{1}^{0}\right)=0$ and
$L\left(\widetilde{\phi}_{1}^{+}\right)=-1-L\left(N_{\alpha}^{c}\right)$.
Hence $L\left(N_{\alpha}^{c}\right)=-1$ and therefore
$L\left(\psi_{\ell,\alpha}\right)=\left(1,1,-1\right)^{T}$ and
$L\left(\phi_{1}\right)=\left(0,0,-2\right)^{T}$. Moving along to the
quark sector, we do not know the lepton number of the third component
of the multiplets $Q_{1,2}$ and $Q_{3}$ (which we call $D_{1,2}$ and
$U$ respectively), but these can be inferred from the interactions
$Q_{1,2}u^{c}\phi_{1}$ and $Q_{3}d^{c}\phi_{1}^{*}$.  Indeed, from the
first interaction it follows that $L\left(D_{1,2}\right)=2$, while the
second one yields $L\left(U\right)=-2$. At this point, the only
$U(1)_{L}$ fermion/scalar charges yet to be found are those of the
components of the scalar triplets $\phi_{2,3}$. But from the
interactions $Q_{1,2}d^{c}\phi_{2,3}$ one readily obtains that
$L\left(\phi_{2,3}\right)=\left(0,0,-2\right)^{T}$.  Note that the
extra Yukawa coupling $Q_{3}u^{c}\phi_{2,3}^{*}$ does preserve this
lepton number assignment.

It is clear that the constrains on the $U(1)_{L}$ charges discussed
above form a linear system of equations, which can be solved at once:
\begin{gather}
	\begin{array}{c}
		\psi_{\ell}\psi_{\ell}\phi_{1}\left\{ \begin{array}{c}
			\\
			\\
			\\
		\end{array}\right.\\
		\psi_{\ell}\ell^{c}\phi_{1}^{*}\left\{ \begin{array}{c}
			\\
			\\
			\\
		\end{array}\right.\\
		Q_{1,2}u^{c}\phi_{1}\left\{ \begin{array}{c}
			\\
			\\
			\\
		\end{array}\right.\hspace{8bp}\\
		Q_{3}d^{c}\phi_{1}^{*}\left\{ \begin{array}{c}
			\\
			\\
			\\
		\end{array}\right.\\
		Q_{1,2}d^{c}\phi_{2,3}\left\{ \begin{array}{c}
			\\
			\\
			\\
		\end{array}\right.\hspace{12bp}\\
		Q_{3}u^{c}\phi_{2,3}^{*}\left\{ \begin{array}{c}
			\\
			\\
			\\
		\end{array}\right.\hspace{8bp}
	\end{array}\hspace{-30bp}\left(\begin{array}{ccccccccc}
	0 & 0 & 0 & 0 & 0 & 1 & 0 & 0 & 0\\
	1 & 0 & 0 & 0 & 1 & 0 & 0 & 0 & 0\\
	1 & 0 & 0 & 1 & 0 & 0 & 0 & 0 & 0\\
	0 & 0 & 0 & 1 & 0 & 0 & 0 & 0 & 0\\
	0 & 0 & 0 & 0 & 1 & 0 & 0 & 0 & 0\\
	1 & 0 & 0 & 0 & 0 & -1 & 0 & 0 & 0\\
	0 & 0 & 0 & 1 & 0 & 0 & 0 & 0 & 0\\
	0 & 0 & 0 & 0 & 1 & 0 & 0 & 0 & 0\\
	0 & 1 & 0 & 0 & 0 & 1 & 0 & 0 & 0\\
	0 & 0 & 0 & 1 & 0 & 0 & 0 & 0 & 0\\
	0 & 0 & 0 & 0 & 1 & 0 & 0 & 0 & 0\\
	0 & 0 & 1 & 0 & 0 & -1 & 0 & 0 & 0\\
	0 & 0 & 0 & 0 & 0 & 0 & 1 & 0 & 0\\
	0 & 0 & 0 & 0 & 0 & 0 & 0 & 1 & 0\\
	0 & 1 & 0 & 0 & 0 & 0 & 0 & 0 & 1\\
	0 & 0 & 0 & 0 & 0 & 0 & 1 & 0 & 0\\
	0 & 0 & 0 & 0 & 0 & 0 & 0 & 1 & 0\\
	0 & 0 & 1 & 0 & 0 & 0 & 0 & 0 & -1
\end{array}\right)\cdot\left(\begin{array}{c}
L\left(N_{\alpha}^{c}\right)\\
L\left(D_{1,2}\right)\\
L\left(U\right)\\
L\left(\phi_{1}^{+}\right)\\
L\left(\phi_{1}^{0}\right)\\
L\left(\widetilde{\phi}_{1}^{+}\right)\\
L\left(\phi_{2,3}^{0}\right)\\
L\left(\phi_{2,3}^{-}\right)\\
L\left(\widetilde{\phi}_{2,3}^{0}\right)
\end{array}\right)=\begin{pmatrix}-2\\
-1\\
-1\\
0\\
0\\
1\\
0\\
0\\
0\\
0\\
0\\
0\\
0\\
0\\
0\\
0\\
0\\
0
\end{pmatrix}\,.
\end{gather}
\thinmuskip=3mu
\medmuskip=4mu plus 2mu minus 4mu
\thickmuskip=5mu plus 5mu
\setlength\tabcolsep{4pt}

We now turn to gauge bosons. The $SU(3)_{L}$ gauge interactions for
triplets $T=\left(T_{1},T_{2},T_{3}\right)^{T}$ and anti-triplets
$A=\left(A_{1},A_{2},A_{3}\right)^{T}$ are of the forms
$-ig_{L}\overline{T}\gamma^{\mu}\mathcal{M}_{\mu}T$ and
$ig_{L}\overline{A}\gamma^{\mu}\mathcal{M}_{\mu}^{T}A$ with
\begin{gather}
	\mathcal{M}_{\mu}=\frac{1}{2}\left(\begin{array}{ccc}
		W_{L,\mu}^{3}+\frac{W_{L,\mu}^{8}}{\sqrt{3}} & W_{L,\mu}^{1}-iW_{L,\mu}^{2} & W_{L,\mu}^{4}-iW_{L,\mu}^{5}\\
		W_{L,\mu}^{1}+iW_{L,\mu}^{2} & \frac{W_{L,\mu}^{8}}{\sqrt{3}}-W_{L,\mu}^{3} & W_{L,\mu}^{6}-iW_{L,\mu}^{7}\\
		W_{L,\mu}^{4}+iW_{L,\mu}^{5} & W_{L,\mu}^{6}+iW_{L,\mu}^{7} & -\frac{2W_{L,\mu}^{8}}{\sqrt{3}}
	\end{array}\right)\,.
\end{gather}
So, given that the lepton number of the components of triplets and
anti-triplets are always of the form of either $\left(x,x,x-2\right)$ or
$\left(y,y,y+2\right)$ for some arbitrary values of $x$ and $y$, it is
clear that gauge interactions preserve the $U(1)_{L}$ we have been
discussing, with $L\left(W_{L,\mu}^{1,2,3,8}\right)=0$ while
$\frac{1}{\sqrt{2}}\left(W_{L,\mu}^{4}\pm iW_{L,\mu}^{5}\right)$ and
$\frac{1}{\sqrt{2}}\left(W_{L,\mu}^{6}\pm iW_{L,\mu}^{7}\right)$
carry $\mp2$ units of lepton number.

One can easily check that with these assignments all terms in the
scalar potential --- except one --- conserve the $U(1)_L$.  This
particular term is identified as $\phi_{1}\phi_{2}\phi_{3}$ and with
the assignments given in table \ref{tab:SVS-reps} it violates $U(1)_{L}$
by two units. If we had switched off the interactions $\psi_{\ell}\psi_{\ell}\phi_{1}$ or
$\psi_{\ell}\ell^{c}\phi_{1}^{*}$ instead of following the procedure above, different $U(1)_L$ symmetries could be defined. Thus, as discussed previously, it is the simultaneous presence of various couplings which violates explicitly lepton number. We remind, however, that even in the 
absence of the trilinear term $\phi_{1}\phi_{2}\phi_{3}$ the SVS 
model does break $U(1)_L$ spontaneously through non-zero
VEVs in the third component of the scalars $\phi_{2,3}$.
\begin{center}
	\begin{figure}[tbph]
		\begin{centering}
			\includegraphics[scale=0.8]{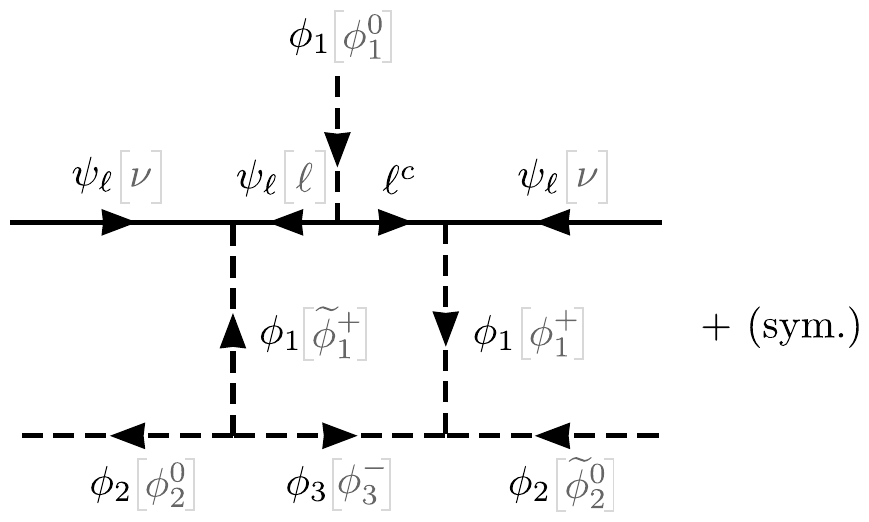} \par\end{centering}
		\protect\caption{\label{fig:SVS_nu_masses}One loop contribution to
			neutrino masses in the original SVS model. There are in total four
			diagrams, since (on top of exchanging the internal $\psi_L$ and
			$\ell^c$) one can exchange everywhere $\phi_{2}$ with $\phi_{3}$. }
	\end{figure}
	\par\end{center}
In the form just presented, the SVS model is not viable as it cannot
accommodate the known neutrino oscillation data. This can be
understood as follows. The $\psi_{\ell}\psi_{\ell}\phi_{1}$
interaction is completely anti-symmetric in the flavour indices. This
leads to the tree-level prediction of a degenerate light neutrino mass
spectrum with eigenvalues $\left(0,m,m\right)$. Since lepton number is
violated in the SVS model, one expects that radiative corrections to
this tree-level result will generate Majorana neutrino masses 
and lead to a non-zero splitting of the degenerate states. 
Figure \ref{fig:SVS_nu_masses} shows an example.
However, in the original SVS model all loop corrections to neutrino
masses are necessarily themselves proportional to the
$\psi_{\ell}\psi_{\ell}\phi_{1}$ interaction, which is the coupling
responsible for the generation of neutrino masses at tree level.
(Indeed, any loop contributing to neutrino masses must
have an odd number of $\psi_{\ell}\psi_{\ell}\phi_{1}$ interactions.) 
The 1-loop corrections are then related to the tree-level mass 
and the relative size of $\delta m_{\nu}^{\rm 1-loop}/m_{\nu}^{\rm tree}$ can 
be estimated to be at most $\sim \frac{1}{16 \pi^2} h_{\tau}^2 \times \cdots < 10^{-6}$, 
where $h_\tau$ is the tau Yukawa coupling and the dots stand for other 
factors which are at most one. 
We will return to a more explicit calculation of this loop in the 
next section. For now it suffices to say that neutrino oscillation 
data requires that the smaller mass splitting in the neutrino 
sector relative to the larger one must be larger than very 
roughly $1/6$, in gross contradiction to the above estimate 
for the original SVS model.

Just for completeness, note that in the diagram of figure
\ref{fig:SVS_nu_masses} the LNV interaction $\phi_{1}\phi_{2}\phi_{3}$
and its conjugate are present, hence the real source of LNV in this
case are the $\widetilde{\phi}_{2}^{0}$ and $\widetilde{\phi}_{3}^{0}$
VEVs. This does not, however, mean that the LNV in the trilinear 
interaction is irrelevant in general. In fact, it is easy to 
built up diagrams containing $\psi_{\ell}\psi_{\ell}\phi_{1}$, 
the SM charged current and this trilinear interaction to generate 
processes such as $e^-e^- \to 4 j$ (or $6j$) at loop-level (tree-level).

\subsection{The Pisano-Pleitez-Frampton (PPF) model}

Following the generic framework of the SVS model, in 1992 a different
331 model was presented \cite{Pisano1992,Frampton1992}. This model
chooses $\beta=-\sqrt{3}$ and thus the third component of the lepton
triplet field $\psi_{\ell}$ has charge +1, hence it is identifiable as
a right-handed charged lepton --- see table \ref{tab:PP-reps}.

\begin{table}[tbph]
	\begin{centering}
		\begin{tabular}{cccccc}
			\toprule 
			Field & 331 representation & $G_{SM}$ decomposition & \# flavours & Components & Lepton number\tabularnewline
			\midrule
			$\psi_{\ell,\alpha}$ & $\left(\mathbf{1},\mathbf{3},0\right)$ & $\left(\mathbf{1},\widehat{\mathbf{2}},-\frac{1}{2}\right)+\left(\mathbf{1},\widehat{\mathbf{1}},1\right)$ & 3 & $\left(\left(\nu_{\alpha},\ell_{\alpha}\right),\ell_{\alpha}^{c}\right)^{T}$ & $\left(1,1,-1\right)^{T}$\tabularnewline
			$Q_{\alpha=1,2}$ & $\left(\mathbf{3},\overline{\mathbf{3}},-\frac{1}{3}\right)$ & $\left(\mathbf{3},\widehat{\mathbf{2}},\frac{1}{6}\right)+\left(\mathbf{3},\widehat{\mathbf{1}},-\frac{4}{3}\right)$ & 2 & $\left(\left(d_{\alpha},-u_{\alpha}\right),J_{\alpha}^{c}\right)^{T}$ & $\left(0,0,2\right)^{T}$\tabularnewline
			$Q_{3}$ & $\left(\mathbf{3},\mathbf{3},\frac{2}{3}\right)$ & $\left(\mathbf{3},\widehat{\mathbf{2}},\frac{1}{6}\right)+\left(\mathbf{3},\widehat{\mathbf{1}},\frac{5}{3}\right)$ & 1 & $\left(\left(t,b\right),J_{3}^{c}\right)^{T}$ & $\left(0,0,-2\right)^{T}$\tabularnewline
			$u_{\alpha}^{c}$ & $\left(\overline{\mathbf{3}},\mathbf{1},-\frac{2}{3}\right)$ & $\left(\overline{\mathbf{3}},\widehat{\mathbf{1}},-\frac{2}{3}\right)$ & 3 & $u_{\alpha}^{c}$ & 0\tabularnewline
			$d_{\alpha}^{c}$ & $\left(\overline{\mathbf{3}},\mathbf{1},\frac{1}{3}\right)$ & $\left(\overline{\mathbf{3}},\widehat{\mathbf{1}},\frac{1}{3}\right)$ & 3 & $d_{\alpha}^{c}$ & 0\tabularnewline
			$J_{\alpha=1,2}$ & $\left(\overline{\mathbf{3}},\mathbf{1},\frac{4}{3}\right)$ & $\left(\overline{\mathbf{3}},\widehat{\mathbf{1}},\frac{4}{3}\right)$ & 2 & $J_{\alpha}$ & -2\tabularnewline
			$J_{3}$ & $\left(\overline{\mathbf{3}},\mathbf{1},-\frac{5}{3}\right)$ & $\left(\overline{\mathbf{3}},\widehat{\mathbf{1}},-\frac{5}{3}\right)$ & 1 & $J_{3}$ & 2\tabularnewline
			$\phi_{1}$ & $\left(\mathbf{1},\mathbf{3},1\right)$ & $\left(\mathbf{1},\widehat{\mathbf{2}},\frac{1}{2}\right)+\left(\mathbf{1},\widehat{\mathbf{1}},2\right)$ & 1 & $\left(\left(\phi_{1}^{+},\phi_{1}^{0}\right),\widetilde{\phi}_{1}^{++}\right)^{T}$ & $\left(0,0,-2\right)^{T}$\tabularnewline
			$\phi_{2}$ & $\left(\mathbf{1},\mathbf{3},-1\right)$ & $\left(\mathbf{1},\widehat{\mathbf{2}},-\frac{3}{2}\right)+\left(\mathbf{1},\widehat{\mathbf{1}},0\right)$ & 1 & $\left(\left(\phi_{2}^{-},\phi_{2}^{--}\right),\widetilde{\phi}_{2}^{0}\right)^{T}$ & $\left(2,2,0\right)^{T}$\tabularnewline
			$\phi_{3}$ & $\left(\mathbf{1},\mathbf{3},0\right)$ & $\left(\mathbf{1},\widehat{\mathbf{2}},-\frac{1}{2}\right)+\left(\mathbf{1},\widehat{\mathbf{1}},1\right)$ & 1 & $\left(\left(\phi_{3}^{0},\phi_{3}^{-}\right),\widetilde{\phi}_{3}^{+}\right)^{T}$ & $\left(0,0,-2\right)^{T}$\tabularnewline
			\bottomrule
		\end{tabular}
		\par\end{centering}
	
	\protect\caption{\label{tab:PP-reps}Field content of the
		Pisano-Pleitez-Frampton (PPF) model \cite{Pisano1992,Frampton1992}. }
\end{table}

A central assertion in \cite{Pisano1992} is that lepton number is
violated by charged scalars and gauge bosons. However, we want to
stress here that this is not the case. Using the procedure outlined
above for the SVS model, the PPF model with the interactions described
in \cite{Pisano1992} preserves the $U(1)_{L}$ symmetry under which
the various fields have the charges indicated in table
\ref{tab:PP-reps}, so there is no explicit lepton number violation in
the model as written down in \cite{Pisano1992}. Moreover, unlike the
SVS model, here all neutral scalar components have $L=0$ hence there
cannot be spontaneous lepton number violation either. Thus the
original model of \cite{Pisano1992} is lepton number conserving. It is
important to note, however, that PPF neglected some quartic scalar
interactions which are allowed by the gauge symmetry. Most notably it
can be shown that the coupling
$\phi_{1}\phi_{2}\phi_{3}^{*}\phi_{3}^{*}$, missing in the original
paper, violates lepton number by two units.

\begin{figure}[tbph]
	\begin{centering}
		\includegraphics[scale=0.8]{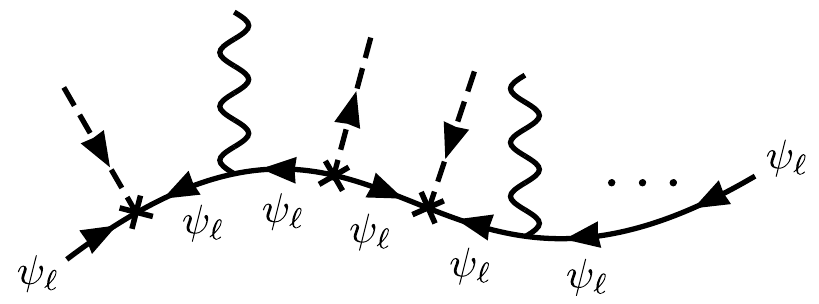}
		\par\end{centering}
	
	\protect\caption{\label{fig:Neutrino_mass_SVS}The only interactions of
		the $\psi_{\ell}$ multiplet in the PPF model are the ones with the
		gauge bosons (which do not change flavour) and those of the form
		$\psi_{\ell}\psi_{\ell}\phi_{3}$, which are anti-symmetric in flavour
		space. Since an odd number of these latter interactions are needed
		to build a mass diagram for $\psi_{\ell}$, such a mass must
		necessarily be flavour anti-symmetric and hence it cannot generate
		(Majorana) neutrino masses.}
\end{figure}

From now on, we will call the version of this model with the most
general gauge invariant Lagrangian the Pisano-Pleitez-Frampton model.
This PPF model is indeed lepton number violating.  Thus, LNV processes
such as neutrinoless double beta decay, will occur. Interestingly, the
PPF model, however, does not generate a non-zero neutrino
mass.\footnote{In the absence of right-handed neutrinos, it would
	necessarily be Majorana-like.}  This can be understood by following
the possible interactions of the $\psi_{\ell}$ triplet, which
contains the SM leptons. Apart from gauge interactions, there is only
the $y_{\ell}\psi_{\ell}\psi_{\ell}\phi_{3}$ Yukawa interaction where
gauge indices are contracted anti-symmetrically. Hence $y_{\ell}$ must
be an anti-symmetric matrix (in flavour space).  Yet, one must
have an odd number of $y_{\ell}$ matrices along the $\psi_{\ell}$
fermion line in any diagram contributing to a neutrino mass matrix
(see figure \ref{fig:Neutrino_mass_SVS}). Hence the flavour matrix
$\mathcal{O}$ associated to the effective operator
$\mathcal{O}_{\alpha\beta}\psi_{\ell,\alpha}\psi_{\ell,\beta}\times (scalars)$ will
always be anti-symmetric (note that the gauge interactions do not
change flavour). Thus, no $\nu\nu$ term will be generated at any order
of a perturbative expansion.

That a LNV model can have zero Majorana neutrino masses but a finite
half-life for neutrinoless double beta decay, seems to be a
contradiction of the well-known ``black-box'' theorem
\cite{Schechter:1981bd}. However, this apparent contradiction can be
traced to another flaw of the PPF model. In it, the
$\psi_{\ell}\psi_{\ell}\phi_{3}$ interaction is the only source of
charged lepton masses. Its antisymmetry implies that the tree-level
charged lepton masses are $(0,m,m)$. (This prediction is analogous to
the one for neutrino masses in the SVS model discussed previously.) This
is in clear disagreement with the experimentally observed charged
lepton masses and thus requires a modification of the PPF model.\footnote{A 
	modified version of the PPF model, which can accommodate
	a realistic charged lepton spectrum, was presented shortly
	after the original one \cite{Foot1993}.  We will come back to this
	in the next section.}  Moreover, this prediction for the charged
lepton spectrum violates the (implicit) assumption in the formulation
of the black box theorem \cite{Schechter:1981bd,Takasugi:1984xr} that
the electron has a non-zero mass. If one follows the procedure given
in the original papers on the black box theorem of completing the
$0\nu\beta\beta$ decay diagram with charged current interactions, in
order to form a Majorana neutrino mass term, one finds that mass
insertions are necessary to convert right-handed electrons into
left-handed ones. For the PPF model all contributions to
$0\nu\beta\beta$ decay produce final states with $e_L e_R$.  The
particular prediction for the charged lepton spectrum in this model
then leads to an
exact zero of the $e_R \to e_L$ insertions, independent of the flavour
compositions of the three mass eigenstates.
This is most easily seen for the case where the only 
non-zero entry in  $y_{\ell}$ is  $y_{\ell}^{\mu\tau}$. In this case, 
the massless state is the electron and it is obvious that  $e_R \to e_L$ 
conversion is impossible. For other cases, the two contributions 
from the degenerate leptons cancel each other exactly. However, 
one expects that once that the PPF model has been modified to 
correct for the unrealistic charged lepton spectrum, non-zero 
Majorana masses will also automatically appear and the standard 
form of the black-box theorem is recovered. A discussion of modified 
PPF models is given below in section \ref{sec:3}.

\subsection{The Pleitez-Özer (PÖ) model}

The generic SVS framework with $\beta=1/\sqrt{3}$ gives rise to the
Pleitez-Özer model \cite{Pleitez:1994pu,Ozer:1995xi}.\footnote{A basic sketch of this model also appears in \cite{Montero:1992jk}.} In it, the third component of
$\psi_{\ell}$ has charge $-1$, so it can be interpreted as the vector
partner of the SM right-handed charged leptons $\ell^{c}$ . Since
there are 3 flavours of $\psi_{\ell}$, 6 copies of $\ell^{c}$ are then
necessary to account for the SM right-handed charged leptons as well
as 3 extra vector fermion pairs $\left(\ell^{c},E\right)$.  There are
no right-handed neutrinos and it can be checked that there is an
unbroken global $U(1)_{L}$ (see table \ref{tab:Ozer-reps}). Furthermore,
none of the neutral scalars carries lepton number, thus
there is also no spontaneous violation of lepton number. Neutrinos are therefore
massless and the model is not satisfactory from this point of view.

\begin{table}[tbph]
	\begin{centering}
		\begin{tabular}{cccccc}
			\toprule 
			Field & 331 representation & $G_{SM}$ decomposition & \# flavours & Components & Lepton number\tabularnewline
			\midrule
			$\psi_{\ell,\alpha}$ & $\left(\mathbf{1},\mathbf{3},-\frac{2}{3}\right)$ & $\left(\mathbf{1},\widehat{\mathbf{2}},-\frac{1}{2}\right)+\left(\mathbf{1},\widehat{\mathbf{1}},-1\right)$ & 3 & $\left(\left(\nu_{\alpha},\ell_{\alpha}\right),E_{\alpha}\right)^{T}$ & $\left(1,1,1\right)^{T}$\tabularnewline
			$\ell_{\alpha}^{c}$ & $\left(\mathbf{1},\mathbf{1},1\right)$ & $\left(\mathbf{1},\widehat{\mathbf{1}},1\right)$ & 6 & $\ell_{\alpha}^{c}$ & -1\tabularnewline
			$Q_{\alpha=1,2}$ & $\left(\mathbf{3},\overline{\mathbf{3}},\frac{1}{3}\right)$ & $\left(\mathbf{3},\widehat{\mathbf{2}},\frac{1}{6}\right)+\left(\mathbf{3},\widehat{\mathbf{1}},\frac{2}{3}\right)$ & 2 & $\left(\left(d_{\alpha},-u_{\alpha}\right),U_{\alpha}\right)^{T}$ & $\left(0,0,0\right)^{T}$\tabularnewline
			$Q_{3}$ & $\left(\mathbf{3},\mathbf{3},0\right)$ & $\left(\mathbf{3},\widehat{\mathbf{2}},\frac{1}{6}\right)+\left(\mathbf{3},\widehat{\mathbf{1}},-\frac{1}{3}\right)$ & 1 & $\left(\left(t,b\right),D\right)^{T}$ & $\left(0,0,0\right)^{T}$\tabularnewline
			$u_{\alpha}^{c}$ & $\left(\overline{\mathbf{3}},\mathbf{1},-\frac{2}{3}\right)$ & $\left(\overline{\mathbf{3}},\widehat{\mathbf{1}},-\frac{2}{3}\right)$ & 5 & $u_{\alpha}^{c}$ & 0\tabularnewline
			$d_{\alpha}^{c}$ & $\left(\overline{\mathbf{3}},\mathbf{1},\frac{1}{3}\right)$ & $\left(\overline{\mathbf{3}},\widehat{\mathbf{1}},\frac{1}{3}\right)$ & 4 & $d_{\alpha}^{c}$ & 0\tabularnewline
			$\phi_{i=1,2}$ & $\left(\mathbf{1},\mathbf{3},\frac{1}{3}\right)$ & $\left(\mathbf{1},\widehat{\mathbf{2}},\frac{1}{2}\right)+\left(\mathbf{1},\widehat{\mathbf{1}},0\right)$ & 2 & $\left(\left(\phi_{i}^{+},\phi_{i}^{0}\right),\widetilde{\phi}_{i}^{0}\right)^{T}$ & $\left(0,0,0\right)^{T}$\tabularnewline
			$\phi_{3}$ & $\left(\mathbf{1},\mathbf{3},-\frac{2}{3}\right)$ & $\left(\mathbf{1},\widehat{\mathbf{2}},-\frac{1}{2}\right)+\left(\mathbf{1},\widehat{\mathbf{1}},-1\right)$ & 1 & $\left(\left(\phi_{3}^{0},\phi_{3}^{-}\right),\widetilde{\phi}_{3}^{-}\right)^{T}$ & $\left(0,0,0\right)^{T}$\tabularnewline
			\bottomrule
		\end{tabular}
		\par\end{centering}
	
	\protect\caption{\label{tab:Ozer-reps}Field content of the Pleitez-Özer model
		\cite{Pleitez:1994pu,Ozer:1995xi}, which conserves lepton number. }
\end{table}

\subsection{Model X}

Finally, in the generic SVS framework it is also possible to have
$\beta=\sqrt{3}$ --- we call this the model X. For this value of
$\beta$, the third component of $\psi_{\ell}$ has charge $-2$. Hence
we need the $\ell_{X}$ representation (charge +2) shown in table
\ref{tab:X-reps} to form a massive, vector fermion pair with this
state once the 331 symmetry is broken. The SM right-handed charged
leptons are then in a separate representation $\ell^{c}$. It is
straightforward to check that this model preserves lepton number, just
like the Pleitez-Özer model. So, in the absence of right-handed neutrinos, it
predicts massless neutrinos.

\begin{table}[tbph]
	\begin{centering}
		\begin{tabular}{cccccc}
			\toprule 
			Field & 331 representation & $G_{SM}$ decomposition & \# flavours & Components & Lepton number\tabularnewline
			\midrule
			$\psi_{\ell,\alpha}$ & $\left(\mathbf{1},\mathbf{3},-1\right)$ & $\left(\mathbf{1},\widehat{\mathbf{2}},-\frac{1}{2}\right)+\left(\mathbf{1},\widehat{\mathbf{1}},-2\right)$ & 3 & $\left(\left(\nu_{\alpha},\ell_{\alpha}\right),\ell_{X,\alpha}^{c}\right)^{T}$ & $\left(1,1,1\right)^{T}$\tabularnewline
			$\ell_{\alpha}^{c}$ & $\left(\mathbf{1},\mathbf{1},1\right)$ & $\left(\mathbf{1},\widehat{\mathbf{1}},1\right)$ & 3 & $\ell_{\alpha}^{c}$ & -1\tabularnewline
			$\ell_{X,\alpha}$ & $\left(\mathbf{1},\mathbf{1},2\right)$ & $\left(\mathbf{1},\widehat{\mathbf{1}},2\right)$ & 3 & $\ell_{X,\alpha}$ & -1\tabularnewline
			$Q_{\alpha=1,2}$ & $\left(\mathbf{3},\overline{\mathbf{3}},\frac{2}{3}\right)$ & $\left(\mathbf{3},\widehat{\mathbf{2}},\frac{1}{6}\right)+\left(\mathbf{3},\widehat{\mathbf{1}},\frac{5}{3}\right)$ & 2 & $\left(\left(d_{\alpha},-u_{\alpha}\right),J_{\alpha}^{c}\right)^{T}$ & $\left(0,0,0\right)^{T}$\tabularnewline
			$Q_{3}$ & $\left(\mathbf{3},\mathbf{3},-\frac{1}{3}\right)$ & $\left(\mathbf{3},\widehat{\mathbf{2}},\frac{1}{6}\right)+\left(\mathbf{3},\widehat{\mathbf{1}},-\frac{4}{3}\right)$ & 1 & $\left(\left(t,b\right),J_{3}^{c}\right)^{T}$ & $\left(0,0,0\right)^{T}$\tabularnewline
			$u_{\alpha}^{c}$ & $\left(\overline{\mathbf{3}},\mathbf{1},-\frac{2}{3}\right)$ & $\left(\overline{\mathbf{3}},\widehat{\mathbf{1}},-\frac{2}{3}\right)$ & 3 & $u_{\alpha}^{c}$ & 0\tabularnewline
			$d_{\alpha}^{c}$ & $\left(\overline{\mathbf{3}},\mathbf{1},\frac{1}{3}\right)$ & $\left(\overline{\mathbf{3}},\widehat{\mathbf{1}},\frac{1}{3}\right)$ & 3 & $d_{\alpha}^{c}$ & 0\tabularnewline
			$J_{\alpha=1,2}$ & $\left(\overline{\mathbf{3}},\mathbf{1},-\frac{5}{3}\right)$ & $\left(\overline{\mathbf{3}},\widehat{\mathbf{1}},-\frac{5}{3}\right)$ & 2 & $J_{\alpha}$ & 0\tabularnewline
			$J_{3}$ & $\left(\overline{\mathbf{3}},\mathbf{1},\frac{4}{3}\right)$ & $\left(\overline{\mathbf{3}},\widehat{\mathbf{1}},\frac{4}{3}\right)$ & 1 & $J_{3}$ & 0\tabularnewline
			$\phi_{1}$ & $\left(\mathbf{1},\mathbf{3},0\right)$ & $\left(\mathbf{1},\widehat{\mathbf{2}},\frac{1}{2}\right)+\left(\mathbf{1},\widehat{\mathbf{1}},-1\right)$ & 1 & $\left(\left(\phi_{1}^{+},\phi_{1}^{0}\right),\widetilde{\phi}_{1}^{-}\right)^{T}$ & $\left(0,0,0\right)^{T}$\tabularnewline
			$\phi_{2}$ & $\left(\mathbf{1},\mathbf{3},1\right)$ & $\left(\mathbf{1},\widehat{\mathbf{2}},\frac{3}{2}\right)+\left(\mathbf{1},\widehat{\mathbf{1}},0\right)$ & 1 & $\left(\left(\phi_{2}^{++},\phi_{2}^{+}\right),\widetilde{\phi}_{2}^{0}\right)^{T}$ & $\left(0,0,0\right)^{T}$\tabularnewline
			$\phi_{3}$ & $\left(\mathbf{1},\mathbf{3},-1\right)$ & $\left(\mathbf{1},\widehat{\mathbf{2}},-\frac{1}{2}\right)+\left(\mathbf{1},\widehat{\mathbf{1}},-2\right)$ & 1 & $\left(\left(\phi_{3}^{0},\phi_{3}^{-}\right),\widetilde{\phi}_{3}^{--}\right)^{T}$ & $\left(0,0,0\right)^{T}$\tabularnewline
			\bottomrule
		\end{tabular}
		\par\end{centering}
	
	\protect\caption{\label{tab:X-reps}Field content of the model X, which conserves lepton number. }
\end{table}

\subsection{The flipped model}

All previous four models follow the SVS framework of placing SM lepton
doublets in triplets of $SU(3)_{L}$, while quark doublets are spread
over one triplet and two anti-triplets. In other words, the extended
gauge symmetry discriminates quark families, but not lepton
families. Recently \cite{Fonseca:2016tbn} we proposed a new model
which reverts this scheme: all three quark families are in equal
representations, while lepton families are not. To achieve gauge
anomaly cancellation and acceptable fermion masses, one of the SM
lepton doublets is placed in a sextet of $SU(3)_{L}$, while the rest
of the fermions are in singlets, triplets and anti-triplets.  As for
the scalars, on top of three triplets $\phi_{1,2,3}$, we have
introduced a sextet $S$ which plays an important role in the
generation of lepton masses, through both tree and loop diagrams. The
full field content of the model is reproduced in table
\ref{tab:flipped-reps}. Note that this construction requires
$\beta=1/\sqrt{3}$.

\thinmuskip=0mu
\medmuskip=0mu
\thickmuskip=0mu
\setlength\tabcolsep{2pt}
\begin{table}[tbph]
	\begin{centering}
		\begin{tabular}{cccccc}
			\toprule 
			Field & 331 rep. & $G_{SM}$ decomposition & \# flav. & Components & Lepton number\tabularnewline
			\midrule
			$L_{e}$ & $\left(\mathbf{1},\mathbf{6},-\frac{1}{3}\right)$ & $\begin{array}{c}
			\left(\mathbf{1},\widehat{\mathbf{3}},0\right)+\left(\mathbf{1},\widehat{\mathbf{2}},-\frac{1}{2}\right)\\
			+\left(\mathbf{1},\widehat{\mathbf{1}},-1\right)
			\end{array}$ & 1 & $\left(\begin{array}{ccc}
			\Sigma^{+} & \frac{1}{\sqrt{2}}\Sigma^{0} & \frac{1}{\sqrt{2}}\nu_{e}\\
			\frac{1}{\sqrt{2}}\Sigma^{0} & \Sigma^{-} & \frac{1}{\sqrt{2}}\ell_{e}\\
			\frac{1}{\sqrt{2}}\nu_{e} & \frac{1}{\sqrt{2}}\ell_{e} & E_{e}
			\end{array}\right)$ & $\left(\begin{array}{ccc}
			-1 & -1 & 1\\
			-1 & -1 & 1\\
			1 & 1 & 3
			\end{array}\right)$\tabularnewline
			$L_{\alpha=\mu,\tau}$ & $\left(\mathbf{1},\mathbf{3},-\frac{2}{3}\right)$ & $\left(\mathbf{1},\widehat{\mathbf{2}},-\frac{1}{2}\right)+\left(\mathbf{1},\widehat{\mathbf{1}},-1\right)$ & 2 & $\left(\nu_{\alpha},\ell_{\alpha},E_{\alpha}\right)^{T}$ & $\left(1,1,3\right)^{T}$\tabularnewline
			$\ell_{\alpha}^{c}$ & $\left(\mathbf{1},\mathbf{1},1\right)$ & $\left(\mathbf{1},\widehat{\mathbf{1}},1\right)$ & 6 & $\ell_{\alpha}^{c}$ & $-1$\tabularnewline
			$Q_{\alpha}$ & $\left(\mathbf{3},\overline{\mathbf{3}},\frac{1}{3}\right)$ & $\left(\mathbf{3},\widehat{\mathbf{2}},\frac{1}{6}\right)+\left(\mathbf{3},\widehat{\mathbf{1}},\frac{2}{3}\right)$ & 3 & $\left(d_{\alpha},-u_{\alpha},U_{\alpha}\right)^{T}$ & $\left(0,0,-2\right)^{T}$\tabularnewline
			$u_{\alpha}^{c}$ & $\left(\overline{\mathbf{3}},\mathbf{1},-\frac{2}{3}\right)$ & $\left(\overline{\mathbf{3}},\widehat{\mathbf{1}},-\frac{2}{3}\right)$ & 6 & $u_{\alpha}^{c}$ & $0$\tabularnewline
			$d_{\alpha}^{c}$ & $\left(\overline{\mathbf{3}},\mathbf{1},\frac{1}{3}\right)$ & $\left(\overline{\mathbf{3}},\widehat{\mathbf{1}},\frac{1}{3}\right)$ & 3 & $d_{\alpha}^{c}$ & $0$\tabularnewline
			$\phi_{i=1,2}$ & $\left(\mathbf{1},\mathbf{3},\frac{1}{3}\right)$ & $\left(\mathbf{1},\widehat{\mathbf{2}},\frac{1}{2}\right)+\left(\mathbf{1},\widehat{\mathbf{1}},0\right)$ & 2 & $\left(H_{i}^{+},H_{i}^{0},\sigma_{i}^{0}\right)^{T}$ & $\left(0,0,2\right)^{T}$\tabularnewline
			$\phi_{3}$ & $\left(\mathbf{1},\mathbf{3},-\frac{2}{3}\right)$ & $\left(\mathbf{1},\widehat{\mathbf{2}},-\frac{1}{2}\right)+\left(\mathbf{1},\widehat{\mathbf{1}},-1\right)$ & 1 & $\left(H_{3}^{0},H_{3}^{-},\sigma_{3}^{-}\right)^{T}$ & $\left(0,0,2\right)^{T}$\tabularnewline
			$S$ & $\left(\mathbf{1},\mathbf{6},\frac{2}{3}\right)$ & $\begin{array}{c}
			\left(\mathbf{1},\widehat{\mathbf{3}},1\right)+\left(\mathbf{1},\widehat{\mathbf{2}},\frac{1}{2}\right)\\
			+\left(\mathbf{1},\widehat{\mathbf{1}},0\right)
			\end{array}$ & 1 & $\left(\begin{array}{ccc}
			\Delta^{++} & \frac{1}{\sqrt{2}}\Delta^{+} & \frac{1}{\sqrt{2}}H_{S}^{+}\\
			\frac{1}{\sqrt{2}}\Delta^{+} & \Delta^{0} & \frac{1}{\sqrt{2}}H_{S}^{0}\\
			\frac{1}{\sqrt{2}}H_{S}^{+} & \frac{1}{\sqrt{2}}H_{S}^{0} & \sigma_{S}^{0}
			\end{array}\right)$ & $\left(\begin{array}{ccc}
			-2 & -2 & 0\\
			-2 & -2 & 0\\
			0 & 0 & 2
			\end{array}\right)$\tabularnewline
			\bottomrule
		\end{tabular}
		\par\end{centering}
	
	\protect\caption{\label{tab:flipped-reps}Field content of the flipped
		331 model \cite{Fonseca:2016tbn}. }
\end{table}
\thinmuskip=3mu
\medmuskip=4mu plus 2mu minus 4mu
\thickmuskip=5mu plus 5mu
\setlength\tabcolsep{4pt}

Without making a rigorous fit, we showed in \cite{Fonseca:2016tbn}
that the model is able to reproduce the observed fermion masses and
mixing angles, hence modifications of the model are not mandatory.
Here, neutrinos are Majorana particles and therefore lepton number is
obviously not conserved by the full Lagrangian. However, if one were
to keep only the gauge interactions as well as all the Yukawa
interactions allowed by the gauge symmetry, one finds a preserved
$U(1)_{L}$ lepton number symmetry, with the associated charges shown
in the last column of table \ref{tab:flipped-reps}. As for scalar
couplings, most of them also preserve this $U(1)_{L}$, including
$\phi_{i}^{*}\phi_{j}\phi_{3}S$ and $\phi_{3}\phi_{3}SS$ which are not
self-conjugate. There are only two interactions allowed by the
gauge symmetry which break lepton number (by two units):
$\phi_{1}\phi_{2}\phi_{3}$ and $\phi_{i}\phi_{j}S^{*}$($i,j=1,2$). In
fact, this last one was mentioned in \cite{Fonseca:2016tbn} as being
important to achieve a realistic neutrino mass matrix. Apart from these
two sources of LNV, one also has to consider the VEVs $\left\langle
\sigma_{i}^{0}\right\rangle $, $\left\langle \Delta^{0*}\right\rangle
$ and $\left\langle \sigma_{S}^{0}\right\rangle $ which all break
$U(1)_{L}$ by two units.

\subsection{The $E_{6}$ inspired model}

The $SU(3)\times SU(3)\times U(1)$ is contained in $SU(3)^{3}$ which
in turn is a subgroup of the exceptional $E_{6}$ group. This group has
been used in Grand Unified model building \cite{Gursey:1975ki}. In
these models, fermions are in three copies of the fundamental
representation $\mathbf{27}$ hence, upon breaking the group down to
the $SU(3)\times SU(3)\times U(1)$ subgroup, one ought to obtain a 331
model with family replication in both the lepton and quark
sectors. Apart from a possible flip between the $SU(3)$'s
representations with their anti-representations, the $E_{6}$
fundamental representation branches as follows:
\begin{gather}
	\mathbf{27}\rightarrow\left(\mathbf{1},\mathbf{3},\left[\begin{array}{c}
		2a\\
		-a+b\\
		-a-b
	\end{array}\right]\right)+\left(\overline{\mathbf{3}},\mathbf{1},\left[\begin{array}{c}
	-2a\\
	a-b\\
	a+b
\end{array}\right]\right)+\left(\mathbf{3},\overline{\mathbf{3}},0\right)\,,\label{eq:27_dec}
\end{gather}
where the square brackets indicate in an economical way different states with different
$U(1)_{X}$ charges, while $a$ and $b$ are parameters
describing the linear combination of two $U(1)$'s which form $U(1)_{X}$.
So, in order to place the left-handed quarks in
$\left(\mathbf{3},\overline{\mathbf{3}},0\right)$, this branching rule
implies that one must have $\beta=-1/\sqrt{3}$. In this case,
$\left(\mathbf{3},\overline{\mathbf{3}},0\right)$ will also contain
the $\left(\mathbf{3},\widehat{\mathbf{1}},-\frac{1}{3}\right)$ SM
representation. Hence, from the second term in (\ref{eq:27_dec}) one
must get two $d^{c}$-like states,
$\left(\overline{\mathbf{3}},\widehat{\mathbf{1}},\frac{1}{3}\right)$,
and one $u^{c}$ like state,
$\left(\overline{\mathbf{3}},\widehat{\mathbf{1}},-\frac{2}{3}\right)$.
Since the leptons (first term in (\ref{eq:27_dec})) have the opposite
$X$ charges to these colored states, we shall have two
$\left(\mathbf{1},\mathbf{3},-\frac{1}{3}\right)$ representations plus
one $\left(\mathbf{1},\mathbf{3},\frac{2}{3}\right)$.  Note that any
$E_{6}$ model is anomaly free, hence this list of 331 fields is so
too. Table \ref{tab:E6model-reps} contains the overall picture.
This model clearly dispels the claim that 331 models predict that the
number of generations has to be necessarily equal to the number of
colours.

\begin{table}[tbph]
	\begin{centering}
		\begin{tabular}{cccccc}
			\toprule 
			Field & 331 representation & $G_{SM}$ decomposition & \# flavours & Components & Lepton number\tabularnewline
			\midrule
			$\psi_{\ell,\alpha}$ & $\left(\mathbf{1},\mathbf{3},-\frac{1}{3}\right)$ & $\left(\mathbf{1},\widehat{\mathbf{2}},-\frac{1}{2}\right)+\left(\mathbf{1},\widehat{\mathbf{1}},0\right)$ & 6 & $\left(\left(\nu_{\alpha},\ell_{\alpha}\right),N_{\alpha}^{c}\right)^{T}$ & $\left(1,1,-1\right)^{T}$\tabularnewline
			$\psi_{X,\alpha}$ & $\left(\mathbf{1},\mathbf{3},\frac{2}{3}\right)$ & $\left(\mathbf{1},\widehat{\mathbf{2}},\frac{1}{2}\right)+\left(\mathbf{1},\widehat{\mathbf{1}},1\right)$ & 3 & $\left(\left(E_{X,\alpha},\nu_{X,\alpha}\right),\ell_{\alpha}^{c}\right)^{T}$ & $\left(1,1,-1\right)^{T}$\tabularnewline
			$Q_{\alpha}$ & $\left(\mathbf{3},\overline{\mathbf{3}},0\right)$ & $\left(\mathbf{3},\widehat{\mathbf{2}},\frac{1}{6}\right)+\left(\mathbf{3},\widehat{\mathbf{1}},-\frac{1}{3}\right)$ & 3 & $\left(\left(d_{\alpha},-u_{\alpha}\right),D_{\alpha}\right)^{T}$ & $\left(0,0,2\right)^{T}$\tabularnewline
			$u_{\alpha}^{c}$ & $\left(\overline{\mathbf{3}},\mathbf{1},-\frac{2}{3}\right)$ & $\left(\overline{\mathbf{3}},\widehat{\mathbf{1}},-\frac{2}{3}\right)$ & 3 & $u_{\alpha}^{c}$ & 0\tabularnewline
			$d_{\alpha}^{c}$ & $\left(\overline{\mathbf{3}},\mathbf{1},\frac{1}{3}\right)$ & $\left(\overline{\mathbf{3}},\widehat{\mathbf{1}},\frac{1}{3}\right)$ & 6 & $d_{\alpha}^{c}$ & 0\tabularnewline
			$\phi_{1}$ & $\left(\mathbf{1},\mathbf{3},\frac{2}{3}\right)$ & $\left(\mathbf{1},\widehat{\mathbf{2}},\frac{1}{2}\right)+\left(\mathbf{1},\widehat{\mathbf{1}},1\right)$ & 1 & $\left(\left(\phi_{1}^{+},\phi_{1}^{0}\right),\widetilde{\phi}_{1}^{+}\right)^{T}$ & $\left(0,0,-2\right)^{T}$\tabularnewline
			$\phi_{i=2,3}$ & $\left(\mathbf{1},\mathbf{3},-\frac{1}{3}\right)$ & $\left(\mathbf{1},\widehat{\mathbf{2}},-\frac{1}{2}\right)+\left(\mathbf{1},\widehat{\mathbf{1}},0\right)$ & 2 & $\left(\left(\phi_{i}^{0},\phi_{i}^{-}\right),\widetilde{\phi}_{i}^{0}\right)^{T}$ & $\left(0,0,-2\right)^{T}$\tabularnewline
			\bottomrule
		\end{tabular}
		\par\end{centering}
	
	\protect\caption{\label{tab:E6model-reps}Field content of the $E_{6}$
		inspired 331 model \cite{Sanchez:2001ua}. }
\end{table}

This 331 model was first studied by Sánchez, Ponce and Martinéz in
\cite{Sanchez:2001ua}. They considered a scalar sector with three
triplet fields with the same quantum numbers as the scalars in the SVS model. With
this field content, there are no sources of explicit lepton number
violation, but the electroweak singlet components
$\widetilde{\phi}_{2,3}^{0}$ inside the two $\phi_{2,3}$ triplets do
lead to spontaneous LNV (see table \ref{tab:E6model-reps}).

While we will not do a complete flavour fit of this model to all
experimental data, we shall briefly describe how it is possible to
obtain realistic lepton masses. To start, consider the notation
$\left\langle \phi_{1}\right\rangle =\left(0,k_{1},0\right)^{T}$,
$\left\langle \phi_{2,3}\right\rangle
=\left(k_{2,3},0,n_{2,3}\right)^{T}$ and note that the only allowed
interactions between $\psi_{\ell}$, $\psi_{X}$ and the scalars are
\begin{align}
	\mathscr{L} & =\cdots+y_{\ell\ell}\psi_{\ell}\psi_{\ell}\phi_{1}+y_{\ell X}^{(2)}\psi_{\ell}\psi_{X}\phi_{2}+y_{\ell X}^{(3)}\psi_{\ell}\psi_{X}\phi_{3}+\textrm{h.c.}+\cdots\,.
\end{align}
Here, $y_{\ell\ell}$ stands for a square $6\times6$ matrix, while
$y_{\ell X}^{(i)}$ are two rectangular $6\times3$ matrices. Hence,
considering only the colorless leptons, 
\begin{gather*}
	\left\langle \mathscr{L}\right\rangle _{\textrm{lepton mass}}=m_{\ell,\alpha\beta}\Psi_{\alpha}^{\ell}\Psi_{\beta}^{\ell^{c}}+m_{\nu,\alpha\beta}\Psi_{\alpha}^{\nu}\Psi_{\beta}^{\nu}
\end{gather*}
with
\begin{gather}
	M_{\nu}=2\left(\begin{array}{ccc}
		0 & k_{1}y_{\ell\ell} & -n_{2}y_{\ell X}^{(2)}-n_{3}y_{\ell X}^{(3)}\\
		\cdot & 0 & k_{2}y_{\ell X}^{(2)}+k_{3}y_{\ell X}^{(3)}\\
		\cdot & \cdot & 0
	\end{array}\right)\,,\\
	M_{\ell}=\left(\begin{array}{cc}
		n_{2}y_{\ell X}^{(2)}+n_{3}y_{\ell X}^{(3)} & ,-k_{2}y_{\ell X}^{(2)}-k_{3}y_{\ell X}^{(3)}\end{array}\right)\,,
\end{gather}
in the basis
$\Psi^{\nu}=\left(\nu_{\alpha},N_{\alpha}^{c},\nu_{X,\beta}\right)^{T}$,
$\Psi^{\ell}=\left(\ell_{\alpha}\right)$ and
$\Psi^{\ell^{c}}=\left(E_{X,\beta},\ell_{\beta}^{c}\right)^{T}$
($\alpha=1,\cdots,6$; $\beta=1,\cdots,3$).

A careful analysis of the neutrino mass matrix reveals that, baring
the existence of special alignments and/or cancellations, one expects
the following mass eigenstates:
\begin{itemize}
	\item Three light Majorana neutrino states $\nu_{M}$ with seesaw masses
	$\mathcal{O}\left(y_{\ell\ell}k_{1}k_{i}/n_{i}\right)$, $i=2,3$,
	and composed almost entirely of $N^{c}$ states;
	\item Three quasi-Dirac neutrino pairs $\nu_{lQD}$ with masses $\mathcal{O}\left(y_{\ell\ell}k_{1}\right)$
	composed of a \textasciitilde{}50\%/50\% admixture of $N^{c}$ and
	$\nu$ states;
	\item Three quasi-Dirac heavy neutrino pairs $\nu_{hQD}$ with masses $\mathcal{O}\left(y_{\ell X}^{(i)}n_{i}\right)$
	composed of a \textasciitilde{}50\%/50\% admixture of $\nu$ and $\nu_{X}$
	states.
\end{itemize}
This rough estimation holds true only if there is a clear hierarchy
between these sets of neutrino masses: $y_{\ell X}^{(i)}n_{i}\gg
y_{\ell\ell}k_{1}\gg y_{\ell\ell}k_{1}k_{i}/n_{i}$ ($i=2,3$). However,
in this limit the three seesawed neutrinos are mostly singlets under
the SM gauge group, hence they cannot play the role of the observed
active neutrinos. That role must then be played by the three
quasi-Dirac neutrino pairs $\nu_{lQD}$ with masses proportional to the
value of the coupling matrix $y_{\ell\ell}$ and the VEV $k_{1}$. Even
though we will not write down the precise expressions for the neutrino
masses and lepton mixing angles, it is possible to have sub-eV active
neutrinos $\nu_{lQD}$, at the price of choosing small 
$\mathcal{O}\left(10^{-12}\right)$ entries in the matrix $y_{\ell\ell}$. This choice does
not affect the mass of the remaining active neutrinos $\nu_{hQD}$,
which must have masses above the SM $Z^0$ mass, in order not to be in
conflict with the measured invisible width of the $Z^0$ boson.  One
must then additionally ensure that the light Majorana neutrino states
$\nu_{M}$ do not mix significantly with the $\nu_{lQD}$ states.

As for charged leptons, the $M_{\ell}$ matrix will have rank 3 if
the matrix $n_{2}y_{\ell X}^{(2)}+n_{3}y_{\ell X}^{(3)}$ is proportional
to $k_{2}y_{\ell X}^{(2)}+k_{3}y_{\ell X}^{(3)}$, and this is an
interesting limit as it would imply that 3 of the charged leptons
are massless ($e$, $\mu$, $\tau$), so a small departure from this
scenario can actually be used to explain the ratio $m_{\tau}/m_{W,Z}$.
Finally, since the quark Yukawa coupling matrices are free parameters,
the quark masses and mixing parameters can easily be fitted in this
model. Since the $E_6$-inspired model can, in principle, explain the 
observed fermion masses, we will not discuss extended versions of 
this 331 model.

\section{\label{sec:3}Simple extensions of the SVS, PPF, P\"O and X models}

Four of the basic 331 models discussed above fail to produce a viable
neutrino mass spectrum. These four follow the basic framework of the
SVS model and we called them SVS, PPF, P\"O and X in the previous
section.  The PPF model moreover predicts a charged lepton spectrum in
disagreement with experimental data, see table
\ref{tab:Problems_with_the_basic_models} for a summary. The table also
recalls, as discussed above, that lepton number is actually {\em
	conserved} in models P\"O and X.

\begin{table}[tbh]
	\begin{centering}
		\begin{tabular}{ccccc}
			\toprule 
			Issue & SVS & PPF & P\"O & X\tabularnewline
			\midrule
			$U(1)_L$ violation? & \cmark & \cmark & \xmark & \xmark\tabularnewline
			$\nu$ masses? & \cmark & \xmark & \xmark & \xmark\tabularnewline
			Correct $\nu$ masses? & \xmark & \xmark & \xmark & \xmark\tabularnewline
			Correct $\ell$ masses? & \cmark & \xmark & \cmark & \cmark\tabularnewline
			\bottomrule
		\end{tabular}
		\par\end{centering}
	
	\protect\caption{\label{tab:Problems_with_the_basic_models} A summary
		of lepton number violation and problems with the lepton sector in
		four of the basic 331 models discussed in section \ref{sec:2}.}
\end{table}

To fix these problems, in the following we consider 4 simple
extensions of the field content for these basic models:
\begin{itemize}
	\item Add a fermionic particle, $N'^{c}$, singlet under the 331
	symmetry group. 
	\item Add a scalar sextet $S$ such that $\psi_{\ell}\psi_{\ell}S$ provides
	a symmetric contribution to the neutrino and charged leptons mass
	matrix.
	\item Add a vector-like pair $\left(E^{c},E\right)$ of charged leptons.
	\item Add a $\phi_{X}$ triplet scalar field in order to generate the
	interaction $\psi_{\ell}\psi_{\ell}\phi_{X}$.
\end{itemize}
However, not all of these extensions work equally well for all 
models, see table \ref{tab:Generic_solutions}. Here, extensions 
which will fix  the problems with the lepton spectra 
for a particular model are marked with (\cmark), while those 
that do not work are marked with (\xmark). The cases which fail 
can be understood as follows:
\begin{itemize}
	\item Adding right-handed neutrinos to the PPF model leads to the generation of neutrino masses, but it does not fix the charged lepton mass problem.
	\item Both the PPF and SVS models already contain a 
	$\psi_{\ell}\psi_{\ell}\phi_{i}$ interaction, 
	so adding another $\phi_{X}$ scalar does not lead to a qualitative change of these models.
	\item Models P\"O and X already contain vector-like leptons, 
	hence adding another pair is again unhelpful.  
	Furthermore, adding the vector fermions $\left(E^{c},E\right)$ to the SVS
	model is also unsatisfactory.
\end{itemize}

Having said this, we now turn to a detailed discussion of the effects of the model extensions which do work.
\begin{center}
	\begin{table}[tbh]
		\begin{centering}
			\begin{tabular}{ccccc}
				\toprule
				Modification & PPF & SVS & P\"O & X \tabularnewline
				\midrule
				$+N'^{c}$  & \xmark & \cmark & \cmark & \cmark \tabularnewline
				$+S$ & \cmark & \cmark & \cmark & \cmark  \tabularnewline
				$+E^{c},E$ & \cmark & \xmark & \xmark & \xmark  \tabularnewline
				$+\phi_{X}$ & \xmark & \xmark & \cmark & \cmark \tabularnewline
				\bottomrule
			\end{tabular}
			\par\end{centering}
		\protect\caption{\label{tab:Generic_solutions}Simple extensions of the
			four models (PPF, SVS, P\"O, X) which will fix (\cmark) the problems
			with the lepton spectra summarized in table
			\ref{tab:Problems_with_the_basic_models}. Cases that will not work
			are marked with (\xmark). For explanation see text.}
	\end{table}
	
	\par\end{center}

\subsection{Extended PPF models}

We start our discussion with the PPF model. Adding a fermion singlet
$N'^{c}$ to the PPF model allows an interaction term
$\psi_{\ell}N'^{c}\phi_{3}^{*}$ (and a mass term $N'^{c}N'^{c}$). 
Since this addition does not affect the charged lepton spectrum, 
by itself such extension of the PPF model is insufficient 
and we will thus not discuss it here (but see below for other 
models).

Adding a scalar sextet, on the other hand, provides a valid fix for the
PPF model. Consider $S=\left(\mathbf{1},\mathbf{6},0\right)$:
\begin{equation}\label{eq:sexplet}
	S = 
	\begin{pmatrix}
		\Delta^{0} & \frac{1}{\sqrt{2}}\Delta^{-} & \frac{1}{\sqrt{2}}H^{+} \\
		\frac{1}{\sqrt{2}}\Delta^{-} & \Delta^{--} & \frac{1}{\sqrt{2}}H^{0} \\
		\frac{1}{\sqrt{2}}H^{+} & \frac{1}{\sqrt{2}}H^{0} & \sigma^{++}
	\end{pmatrix}\,.
\end{equation}
The components denoted as $\Delta$, $H$, and $\sigma$ form a triplet, a doublet, and a singlet respectively under the $SU(2)_L$ group. The interaction of the 
lepton triplet with this sextet contains the terms
\begin{equation}\label{eq:sxpint}
	y_S \psi_{\ell,\alpha} S^* \psi_{\ell,\beta} = y_S \left[ \nu_{\alpha}\nu_{\beta}\Delta^{0*}
	+ \frac{1}{\sqrt{2}} (\ell_{\alpha}\ell^{c}_\beta+\ell^{c}_\alpha \ell_{\beta})H^{0*} 
	+ \cdots \right]\,.
\end{equation}
The term proportional to $\Delta^{0*}$ will give a type-II seesaw
contribution to the neutrino masses, once $\Delta^0$ acquires a VEV,
proportional to $m_{\alpha\beta}^{\nu} = (y_{S})_{\alpha\beta} \langle
\Delta^0\rangle$, while the charged lepton mass matrix is now
the sum of two terms: $m^\ell = y_\ell \langle \phi^0_3\rangle + y_S \langle
H^0\rangle$.  It is easy to see that in the absence of $y_S$ the mass
spectrum for the charged leptons has the eigenvalues ($0,m,m$). Thus, 
in order to achieve the correct hierarchies for $e$, $\mu$ and $\tau$, 
the second term in $m^\ell$ must dominate. This puts a lower limit on 
the largest entries in $ y_S \langle H^0\rangle$ of the order of the 
$\tau$ mass. 

Since the same $y_S$ appears in neutrino masses, one must have $\langle\Delta^0\rangle/\langle H^0\rangle \lsim 
10^{-10}$ for a correct explanation of neutrino data.
Adding a sextet to the original PPF model was already proposed in 
\cite{Foot1993}. These authors, however, argued that such a small 
ratio calls for a protecting symmetry. The proposed symmetry 
eliminates all lepton number violating scalar interactions 
from the model: in addition to the original term 
$\phi_1\phi_2\phi_3^*\phi_3^*$, these are $\phi_3 \phi_3 S^*$ and 
$SSS$. Since under this 
condition lepton number is conserved, neutrinos are massless 
again. Thus, with the addition of only a sextet to the original 
PPF model, we have to accept the fine-tuning between the triplet 
and doublet VEVs if we are to explain neutrino data. 
We note in passing that such a small ratio of VEVs might be
due to a small parameter in the scalar potential, such as the 
coefficient of $\phi_3 \phi_3 S^*$.

\begin{center}
	\begin{figure}[tbph]
		\begin{centering}
			\includegraphics[scale=0.72]{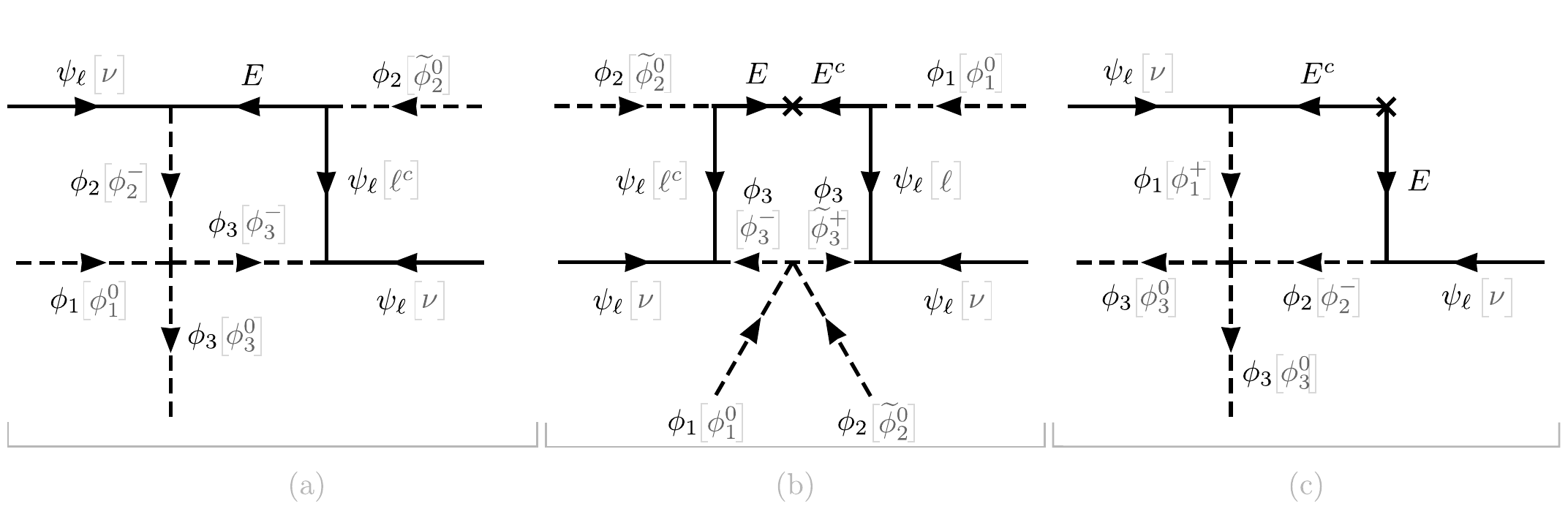}
			\par\end{centering}
		
		\protect\caption{\label{fig:PPFloops}{One-loop diagrams 
				in the PPF model, extended with a vector-like lepton.}}
	\end{figure}
	\par\end{center}

We now turn to the third possibility in our list.  Both problems,
neutrino and charged lepton masses, can be cured in the PPF model by
the introduction of a pair of vector-like charged leptons, $E$ and
$E^c$, in the representations $\left(\mathbf{1},\mathbf{1},\mp
1\right)$.\footnote{That the wrong prediction for the charged lepton 
spectrum in the PPF model can be cured using vector-like leptons 
was noted already in \cite{Duong:1993zn,Montero:2001ji}.}
In the original PPF model, the Lagrangian contains the
interaction terms
\begin{equation}\label{eq:LagPPF}
	{\mathscr{L} = \cdots + \frac{1}{2}y_{\ell,\alpha\beta}\psi_{\ell,\alpha}\psi_{\ell,\beta}\phi_3 
		+ \lambda_7 \phi_1\phi_2\phi_3^*\phi_3^* + \textrm{h.c.}\,,}
\end{equation}
and introducing the left-handed Weyl spinors $E$ and $E^c$ makes it possible to write down the following also
\begin{equation}\label{eq:LagEc}
	\mathscr{L}_{EE^c} = h_{E^c,\alpha} \psi_{\ell,\alpha} E^c \phi_1^* 
	+ h_{E,\alpha} \psi_{\ell,\alpha} {E} \phi_2^* + m_{EE^c} E{E^c}\,.
\end{equation}
The charged lepton mass matrix, after symmetry breaking, becomes:
\begin{equation}\label{eq:MassChLep}
	M_{\ell E} =
	\begin{pmatrix}
		0 & y_{\ell,e\mu}k_3 & y_{\ell,e\tau}k_3 & h_{E^c,e}k_1 \\
		-y_{\ell,e\mu}k_3 & 0 & y_{\ell,\mu\tau}k_3 & h_{E^c,\mu}k_1 \\
		-y_{\ell,e\tau}k_3  & - y_{\ell,\mu\tau}k_3 & 0 & h_{E^c,\tau}k_1 \\
		h_{E,e} n_2 & h_{E,\mu} n_2 & h_{E,\tau} n_2 & m_{EE^c}
	\end{pmatrix} .
\end{equation}
Here, $k_1$, $k_3$ and $n_2$ are the VEVs of $\phi^0_1$, $\phi^0_3$
and ${\widetilde \phi}^0_2$, respectively. We now define $|y_\ell| =
\sqrt{y_{\ell,e\mu}^2+y_{\ell,e\tau}^2+y_{\ell,\mu\tau}^2}$, $|h_{E^c}| =
\sqrt{\sum_\alpha (h_{E^c,\alpha})^2}$ and $|h_{{E}}| = \sqrt{\sum_\alpha
	(h_{{E,\alpha}})^2}$.  Then, in the limit $|h_{E^c}|=|h_{{
		E}}|\to 0$ we obtain the original result for the charged lepton
masses, given by: $m_{1,2,3} = (-|y_\ell|k_3,0,|y_\ell|k_3)$. Note that the
massless state has the eigenvector $e_2 =
\frac{1}{|y_\ell|}(y_{\ell,\mu\tau},y_{\ell,e\tau},1)^T$.  In other words, a good starting point
to have the electron as the lightest state corresponds to the choice
$y_{\ell,e\mu},y_{\ell,e\tau} \ll y_{\ell,\mu\tau}$.  Note that $|y_\ell| \to 0$ is not
allowed, since in this case the matrix $M_{\ell E}$ in equation (\ref{eq:MassChLep}) only has 2
non-zero eigenvalues. In this limit, the lighter of the two non-zero
mass states is given by $m =|h_{{E}}| |h_{{E^c}}|k_1n_2/m_{EE^c}$.  For values of $m_{EE^c}={\cal O}$(TeV) and
$n_2={\cal O}$(TeV), fitting $m_{\tau}$ then requires 
$|h_{E}| \times |h_{E^c}| \sim {\cal   O}(10^{-2})$. For non-zero values
of $|h_{E}|$, $|h_{E^c}|$ and $y_\ell$ the mass degeneracy is
broken and a realistic charged lepton spectrum can be easily obtained
for $|y_\ell| \lsim {\cal O}(10^{-2}-10^{-3})$ together with $|y_\ell| < |h_{E}|
|h_{E^c}|$ (and $k_1 \sim k_3$).

Extending the PPF model with a vector-like lepton does not only solve
the charged lepton mass problem, but it also leads to the generation of
1-loop neutrino masses --- see figure \ref{fig:PPFloops}.\footnote{We show
	the loops with the internal scalars as propagating
	degrees of freedom. In a full calculation one should take into
	account that the same scalars are used to break the 331 symmetry,
	i.e. some components of these scalars become the Goldstone bosons
	and are ``eaten'' by the massive vectors. This will lead to the
	generation of equivalent diagrams, but now with vector bosons. We
	will omit this (irrelevant) complication in our discussion here.}
For these loops, in addition to the terms given in
equation (\ref{eq:LagEc}), the two interactions terms in
equation (\ref{eq:LagPPF}) are needed. As explained in the previous
section, in the minimal PPF model lepton number violation is
proportional to $\lambda_7$ and, thus, all loops that generate a
Majorana neutrino mass must contain this particular quartic vertex. 
Such statement is still true once $E$ and $E^c$ are added to the model, hence
this important scalar interaction is present in all diagrams show in
figure \ref{fig:PPFloops}. 

We will give a rough estimate of the size of these loops. A complete
calculation would require rotating all internal states in the diagrams
to the mass eigenstate basis and then summing over all states. However,
since (i) the mass of the vector-like lepton has to be much larger
than the mass of the tau and (ii) $\lambda_7$ has to be small, as shown
below, we can estimate the relative contributions of each diagram in
figure \ref{fig:PPFloops} to the neutrino mass individually. Let's concentrate on
diagram (c) first. It's contribution to the neutrino mass matrix 
is estimated to be:
\begin{equation}\label{eq:diagC}
	(m_{\nu})_{\alpha\beta}= \frac{1}{16 \pi^2}\sin 2\theta_S m_{EE^c}\Delta B_0
	\left[ h_{E,\alpha}h_{E^c,\beta} + (\alpha \leftrightarrow \beta) \right]\,.
\end{equation}
Here, $\theta_S$ is the angle that diagonalizes the (2,2) submatrix 
of the charged scalars,
\begin{equation}\label{eq:MsqChS}
	M^2_{\phi_1\phi_2} =
	\begin{pmatrix}
		m_{\phi_1}^2 & \lambda_7 k_3^2 \\
		\lambda_7 k_3^2 & m_{\phi_2}^2 
	\end{pmatrix}\,,
\end{equation}
and is given by
\begin{equation}\label{eq:tantheta}
	\sin 2\theta_S = \frac{2 \lambda_7 k_3^2}{m_{1}^2-m_{2}^2}\,,
\end{equation}
with $m_{1,2}^2$ being the eigenvalues of $M^2_{\phi_1\phi_2}$. 
In equation (\ref{eq:diagC}) $\Delta B_0$ stands for the difference between the 
two 1-loop $B_0$ functions for the two scalar mass eigenstates 
and it reads
\begin{equation}\label{eq:DelB}
	\Delta B_0 = \frac{m_1^2  \log(m_1^2/m_{EE^c}^2)}{m_1^2 - m_{EE^c}^2}
	-  \frac{m_2^2  \log(m_2^2/m_{EE^c}^2)}{m_2^2 - m_{EE^c}^2}\,.
\end{equation}
Since we know experimentally that neutrino masses are small, while the
mass of the vector-like lepton should be larger than several 100's
of GeV, either the Yukawa couplings or the factor $\sin 2\theta_S\Delta
B_0$ should be small. The former is not an option in the present
model, since only for $|h_{E}| |h_{{E^c}}| \sim {\cal
	O}(10^{-2})$ a realistic charged lepton spectrum can be obtained, 
as discussed above. 

We define:
\begin{align}
	{\overline M}= & \;\frac{1}{2}\left(m_1+m_2\right)\,, \\
	\Delta M = & \;\left(m_2 -m_1\right)\,.
\end{align}
Then, in the limit of $\Delta M \ll {\overline M}$ and for $m_{EE^c}<{\overline M}$, 
$\sin 2\theta_S\Delta B_0$ becomes simply 
$\sim (\lambda_7 k_3^2)/{\overline M}^2$, so the neutrino mass
is roughly given by the expression
\begin{equation}\label{eq:mnuNum}
	m_{\nu} \sim \left(\frac{m_{E E^c}}{\rm TeV} \right)  
	\left(\frac{\rm TeV}{{\overline M}^2}\right)^2
	\left(\frac{k_3}{100 \hskip1mm {\rm GeV}}\right)^2
	\left(\frac{|h_{E^c}| |h_{{\bar E^c}}| }{10^{-2}}\right)
	\left(\frac{\lambda_7}{10^{-7}}\right) 10^{-1}\hskip 1mm {\rm eV}\,.
\end{equation}
We now turn to a brief discussion of the relative importance of the
diagrams (a)--(c) in figure \ref{fig:PPFloops}. Diagrams (a) and (b)
contain the same parameters as diagram (c) just discussed and,
furthermore, they also depend on the other doublet VEV in the model
($k_1$) as well as the couplings $y_{\ell,\alpha\beta}$. Assuming 
that the Yukawas $h_{E,\alpha}$ and $h_{E^c,\beta}$ are very roughly of 
the same order of magnitude numerically, the relative
importance of the three diagrams can then be estimated to be
\begin{equation}\label{eq:comploop}
	(c) : (a) : (b)  = 1 : |y_\ell|\frac{k_1}{k_3} : \left(|y_\ell|\frac{k_1}{k_3}\right)^2\,.
\end{equation}
The ratio $\frac{k_1}{k_3}$ is not fixed in this model, therefore $k_3$ can be smaller than $k_1$. Only the combination
$\sqrt{k_1^2 + k_3^2}=v=174$ GeV is fixed.
However, as discussed above, the tau mass constrains the combination
$|h_{E}| |h_{E^c}| k_3$ to be of the GeV order, for
$n_2\simeq m_{EE^c}$ of the TeV order. For Yukawa couplings in the perturbative
regime, this means that $k_3$ can not be much smaller than 1 GeV. 
Thus, diagram (c) is usually the dominant one, and the other 
diagrams can be at most equally important, if $k_3$ is pushed to 
its lower limit.

Before moving on, we recall that adding additional triplet scalars,
without adding ($E,E^c$), does not provide a valid solution for the
PPF model, see table \ref{tab:Generic_solutions}.

\subsection{Extended SVS models}

For the SVS model, two of the four possibilities listed in table 
\ref{tab:Generic_solutions} will be valid solutions: (i) adding 
a fermion singlet $N'^c$ and (ii) adding a scalar sextet.

Adding (three copies) of $N'^c$, in addition to the term 
$y_\ell \psi_\ell \psi_\ell \phi_1$, one can write down three new 
Lagrangian terms for the (extended) SVS model:
\begin{equation}\label{eq:LagSVSe}
	\mathscr{L} = \sum_{j=2,3} y^{(j)}_{N'} \psi_\ell N'^c \phi_j^* + \mu N'^cN'^c\,.
\end{equation}
In the basis ($\nu,N^c,N'^c$), the neutrino mass matrix becomes
\begin{equation}\label{eq:mnuSVSe}
	M^{\nu} =
	\begin{pmatrix}
		0 & m_D & m_L \\ 
		m_D^T & 0 & M_R \\ 
		m_L^T & M_R^T & \mu 
	\end{pmatrix}.
\end{equation}
Here, $m_D = y_\ell k_1$, $m_L = \sum_j y^{(j)}_{N'} k_j$, $M_R = \sum_j
y^{(j)}_{N'} n_j$ and $\mu$ are $3 \times 3$ matrices.\footnote{We keep following here the convention
	that $k_i$($n_i$) is the $SU(2)_L$ doublet(singlet) VEV of the scalar triplet $\phi_i$.}
There are two 
limits for $\mu$. For $M_R \ll \mu$ the matrix in equation (\ref{eq:mnuSVSe}) 
will lead to a double seesaw, in other words, integrating out $N'^c$ would 
give a Majorana mass entry in the (2,2) position of the above 
matrix of the order of $M_R \mu^{-1} M_R^T$. If $\mu \ll M_R$, the
matrix gives neutrinos a mass via the inverse seesaw mechanism
\begin{equation}\label{eq:InvSS}
	m_{\nu} = m_D (M_R^T)^{-1} \mu M_R^{-1} m_D^T\,.
\end{equation}
The fit to neutrino masses can easily be done. This case has been
studied in \cite{Boucenna:2015zwa}. Note that if $\mu=0$ there is no linear seesaw contribution proportional to $m_L$. Indeed, in a model such as this one where $m_L\propto M_R$ and $ m_D^T=- m_D$, one has:
\begin{equation}\label{eq:LinSS}
	m_D (M_R^T)^{-1}m_L + \left[m_D (M_R^T)^{-1}m_L\right]^T = 0\,.
\end{equation}

Such limit $\mu=0$ can be achieved with some additional symmetry, as discussed in \cite{Boucenna2014}. However, neutrinos will still acquire mass at 1-loop level via,
for example, the diagram shown in figure \ref{fig:SVS_nu_masses}, and
also via the gauge loops discussed in \cite{Boucenna2014}. Consider first the loop shown in figure \ref{fig:SVS_nu_masses}.  The
loop will vanish in the limit where the coefficient $\rho$ of the term
$\phi_1\phi_2\phi_3$ vanishes.  The calculation is very similar to the
loop discussed for the PPF model, with some modifications: $m_{EE^c}$
has to be replaced by the SM charged lepton masses and the Yukawa
matrices appearing at the vertices are $y_\ell$ and $y_{\ell\ell^c}$, where the
latter is the matrix entering the charged lepton mass matrix.  
If $\rho/{\overline M}$ is a small number, where ${\overline M}$ is some
average mass of the scalars, we very roughly estimate that
\begin{align}\label{eq:EstLpSVS}
	m_{\nu} & \; \sim \frac{2}{16 \pi^2} m_{\tau} 
	\frac{k}{n}\left(\frac{\rho}{{\overline M}}\right)^2
	\left(\frac{n}{{\overline M}}\right)^2 |y_\ell| |y_{\ell\ell^c}|\\
	& \; \sim 0.05 \Big(\frac{\rho/{\overline M}}{10^{-2}}\Big)^2
	\frac{|y_\ell|}{10^{-2}}
	\frac{|y_{\ell\ell^c}|}{10^{-2}} \hskip1mm {\rm eV}\,
\end{align}
for $k_i \simeq k \sim 100$ GeV and $n_i \simeq n \sim 1$ TeV. 

\begin{center}
	\begin{figure}[tbph]
		\begin{centering}
			\includegraphics[scale=0.85]{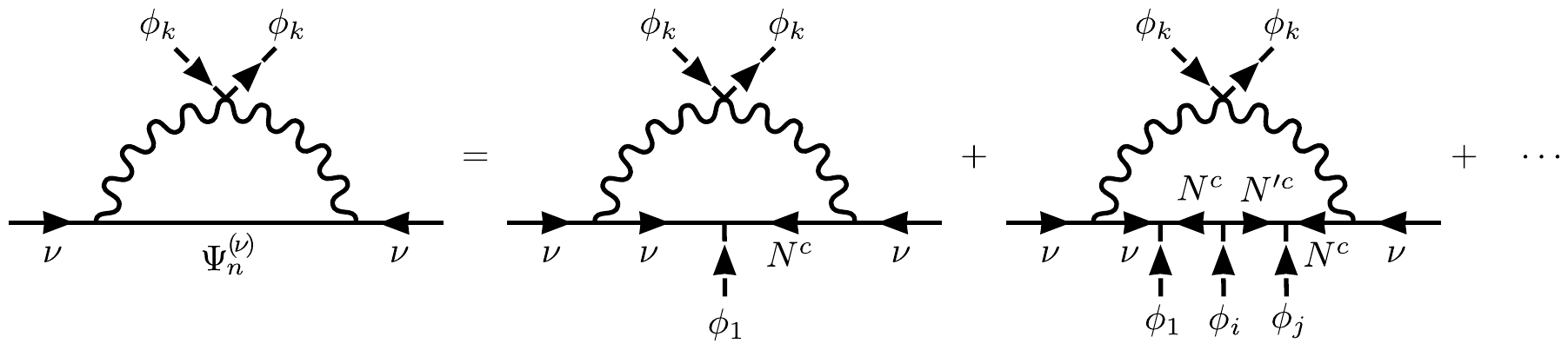}
			\par\end{centering}
		
		\protect\caption{\label{fig:BMVloops}Gauge loops in the extended 
			SVS model. The full neutrino propagator with the mass eigenstates $\Psi^{(\nu)}_n$ can be expanded in a series of mass insertions. The first non-zero term involves three mass insertions (see text).}
	\end{figure}
\end{center}

The gauge loops discussed in \cite{Boucenna2014} are more subtle. 
In the SVS model, in addition to the trilinear coupling $\rho$, 
the VEVs $n_2$ and $n_3$ also violate lepton number. Thus, once the 
331 symmetry is broken, there exists a mixing between gauge 
bosons that leads to lepton number violating processes. In 
particular, one can draw the diagrams shown in figure 
\ref{fig:BMVloops}. Note that the VEV insertions indicated at 
the top of these diagrams always are in the combination 
$k_2 n_2$ and/or $k_3 n_3$, i.e. they correspond to a $\Delta(L)=2$ 
effect. 

The diagram on the left shows the contribution to the neutrino mass in
the basis where the internal fermions are mass eigenstates.  One can
understand this propagator as an infinite series of mass insertions, 
as indicated by the diagrams to the
right. The first term in this expansion is proportional to $m_D$,
which is completely antisymmetric, and thus does not give any
contribution. However, higher order terms will come proportional to
powers of $f=(m_DM_R^*M_R - M_R M_R^{\dagger}m_D)$, which in general
is non-zero. It is interesting to note that, for the
special case where the heavy Dirac-pairs start out degenerate ($M_R\propto \mathbb{1} $), 
the commutator $f$ vanishes and the gauge loops go to zero. 
In the general case, where $f$ is not much smaller than $M_R M_R^{\dagger}m_D$ 
this gauge loop will dominate over the scalar loop and put a 
constraint on $m_D M_R^{-1}$ to be typically below $10^{-8}$ or so.

Adding a sextet $S$ with the quantum numbers $\left(\mathbf{1},\mathbf{6},2/3\right)$ also may solve the neutrino mass problem. The components of such a field can be written as in equation (\ref{eq:sexplet}), with the only difference being the electric charges. The part of the Lagrangian involving $S$ contains the following important terms:
\begin{equation}\label{eq:6svs}
	\mathscr{L} = \cdots + y_S (\Delta^0 \nu \nu + \sqrt{2}H^0 \nu N^c 
	+ \sigma^0 N^c N^c) + \textrm{h.c.}\,.
\end{equation}
If all the VEVs of $\Delta^0$, $H^0$ and $\sigma^0$ are non-zero, the
light neutrino masses have both seesaw type-I and type-II 
contributions. One just needs to ensure that $\langle \Delta^0 \rangle \sim\textrm{eV} \ll
\langle H^0 \rangle \sim\textrm{100 GeV} \ll \langle \sigma^0\rangle \sim\textrm{TeV}$.

\subsection{Extending the P\"O and X models}

The situation is rather simpler in models P\"O and X, which both conserve
lepton number. They also do not have neutrino singlets, hence they predict
that neutrinos are massless.
Here, we will very briefly discuss the different extended versions of
these models, commenting also on the differences with respect to the
models SVS and PPF.  Since models P\"O and X are very similar in this respect,
we discuss both at the same time.

Adding three copies of fermion singlets $N^c$ makes it possible to write down 
the terms
\begin{equation}\label{eq:SSox}
	\mathscr{L} = y_{\nu}\psi_\ell N^c \phi_3^* + M_N N^c N^c\,.
\end{equation}
Note that, since the triplet $\psi_{\ell}$ does not contain 
a $N^c$ in neither model P\"O nor model X, this will give an ordinary 
seesaw mechanism of type-I (to be compared with the inverse or double seesaw 
in the SVS model) which is sufficient to explain neutrino data.

Adding a sextet $S$, with the quantum numbers 
$\left(\mathbf{1},\mathbf{6},-4/3\right)$ in the case of model P\"O and
$\left(\mathbf{1},\mathbf{6},2\right)$ 
in case of model X, gives rise to Majorana neutrino masses once 
the neutral component $\Delta^0$ of the $SU(2)_L$ scalar triplet contained
in $S$ acquires a VEV. This is a pure seesaw type-II contribution since
$\Delta^0$ is the only neutral component of these sextets.

\begin{center}
	\begin{figure}[tbph]
		\begin{centering}
			\includegraphics[scale=0.85]{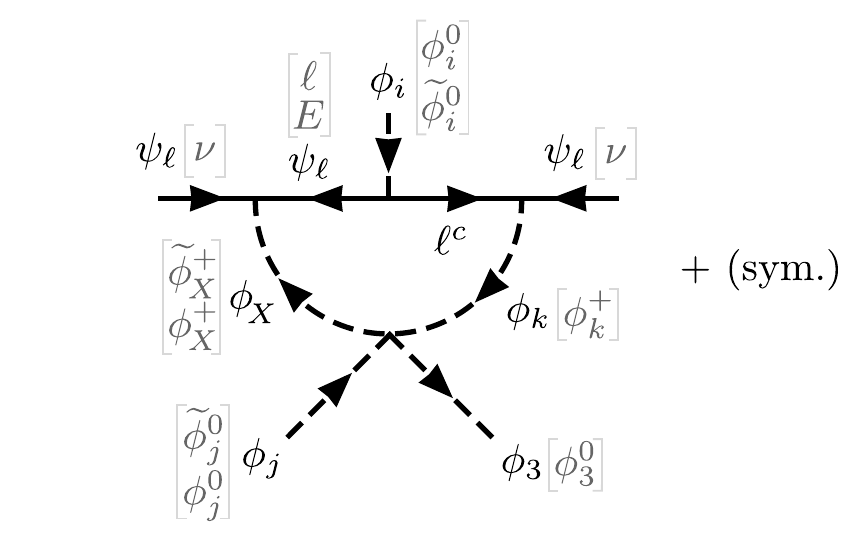}
			\par\end{centering}
		
		\protect\caption{\label{fig:Ozerloop}Scalar loop for neutrino masses in the  
			Pleitez-\"Ozer model extended with a scalar triplet $\phi_X$. In fact, two distinct diagrams are possible, depending on which components inside the square brackets are picked (either the ones on top, or the ones at the bottom). Analogous loops can be made for model X with an added $\phi_X$ scalar triplet.}
	\end{figure}
\end{center}

Finally, neutrino masses can be generated at the 1-loop level also in the models 
P\"O and X, by introducing an additional triplet scalar $\phi_X$. 
The required quantum numbers are $\left(1,\mathbf{3},4/3\right)$ (model P\"O)
and $\left(1,\mathbf{3},2\right)$ (model X). The resulting Feynman diagram, 
in the  Pleitez-\"Ozer model, is shown in figure \ref{fig:Ozerloop}. In both
models the calculation of the loop and the resulting constraints on model
parameters are very similar to 
the results discussed above for models PPF and SVS, with some 
obvious replacements.

\section{\label{sec:4}Conclusions}

We have studied in a systematic way the status of lepton number in 331 
models. The fact that lepton number
often does not commute with the extended $SU(3)_{C}\times 
SU(3)_{L}\times U(1)_{X}$ gauge group makes this an interesting topic,
leading to the existence of gauge bosons
and colored fermions with a non-zero $U(1)_L$ charge and, potentially, to
lepton number violation.
Note also that the 331 symmetry may break to the
Standard Model gauge group at a relatively low energy scale ($\sim$TeV), in which case
the LHC would be able to probe the sources of lepton number violation.

However, as we have made clear in this work, there is a large diversity of 331 models,
and in some of them
lepton number not only commutes with the gauge group, but it is also preserved by
the full Lagrangian and VEVs of the scalars. 
These are nevertheless exceptional cases; in general it is possible to
(a) write down sets of gauge invariant interactions which do not preserve any
global $U(1)_L$ and/or (b) have neutral scalar components with a non-zero
lepton number which break spontaneously this symmetry.

Most of the models we discuss, in their original form, are unable to explain
the observed lepton masses and  neutrino oscillation data.  For these models
we have listed several simple extensions which can accommodate all 
lepton data (some of them had already been proposed previously by other authors).
As such, any of these extended models can be used for further study.

We have focused mainly on the generation of acceptable neutrino masses
(and mixing angles), having mentioned lepton number violating processes,
such as neutrinoless double beta decay, only in passing when it was most relevant.
Elsewhere \cite{Fonseca:2016temp}, we shall provide a more detailed analysis of 
this process, both in 331 models as well as in other models 
with an extended gauge groups.

\section*{\vspace{-0.15cm}Acknowledgments}

This work was supported by the Spanish grants FPA2014-58183-P, Multidark
CSD2009-00064 and SEV-2014-0398 (from the \textit{Ministerio de Economía
	y Competitividad}), as well as PROMETEOII/2014/084 (from the \textit{Generalitat
	Valenciana}).

\clearpage

\appendix

\section*{\label{sec:Appendix-A}Appendix: decomposition of 331 representations\addcontentsline{toc}{section}{\protect\numberline{}Appendix A: contractions and decomposition of 331 representations}}

The decomposition of the most relevant $SU(3)_{L}\times U(1)_{X}$
representations into $SU(2)_{L}\times U(1)_{Y}$ representations has
already been provided in equations (\ref{eq:triplet_decomposition})--(\ref{eq:octet_decomposition}).
As such, in this appendix we simply clarify how the components of
these representations are related.

A triplet $\mathbf{3}$ of $SU(3)_{L}$ breaks into a doublet $\widehat{\mathbf{2}}$
plus a singlet $\widehat{\mathbf{1}}$ of $SU(2)_{L}$. Noting that
the electric charge of each component depends on the $U(1)_{X}$ charge
of the triplet, as well as the $\beta$ parameter as shown in equation
(\ref{eq:triplet_decomposition}), we may simply label the components
of $\widehat{\mathbf{2}}$ by their isopin ($1/2$ and $-1/2$). We
can then settle with the following identification:
\begin{align}
	\mathbf{3} & =\left(\begin{array}{c}
		\widehat{\mathbf{2}}_{1/2}\\
		\widehat{\mathbf{2}}_{-1/2}\\
		\widehat{\mathbf{1}}
	\end{array}\right)\,.
\end{align}
From here we infer that an anti-triplet $\overline{\mathbf{3}}$ of
$SU(3)_{L}$, which also decomposes into a doublet $\widehat{\mathbf{2}}$
plus a singlet $\widehat{\mathbf{1}}$ must be written as
\begin{align}
	\overline{\mathbf{3}} & =\left(\begin{array}{c}
		\widehat{\mathbf{2}}_{-1/2}\\
		-\widehat{\mathbf{2}}_{1/2}\\
		\widehat{\mathbf{1}}
	\end{array}\right)\,.
\end{align}

A (anti)sextuplet $\mathbf{6}$ of $SU(3)_{L}$ breaks into a triplet
$\widehat{\mathbf{3}}$, a doublet $\widehat{\mathbf{2}}$ and a singlet
$\widehat{\mathbf{1}}$ of $SU(2)_{L}$. These representations ($\mathbf{6}$
and $\overline{\mathbf{6}}$) are often pictured as matrices instead
of vectors, since that makes their contraction with triplets more
intuitive. For example, if $\overline{\mathbf{6}}_{ij}\mathbf{3}{}_{i}\mathbf{3}'_{j}$
is gauge invariant, one must have the following identification:
\begin{align}
	\mathbf{6}=\left(\begin{array}{ccc}
		\widehat{\mathbf{3}}_{1} & \frac{1}{\sqrt{2}}\widehat{\mathbf{3}}_{0} & \frac{1}{\sqrt{2}}\widehat{\mathbf{2}}_{1/2}\\
		\frac{1}{\sqrt{2}}\widehat{\mathbf{3}}_{0} & \widehat{\mathbf{3}}_{-1} & \frac{1}{\sqrt{2}}\widehat{\mathbf{2}}_{-1/2}\\
		\frac{1}{\sqrt{2}}\widehat{\mathbf{2}}_{1/2} & \frac{1}{\sqrt{2}}\widehat{\mathbf{2}}_{-1/2} & \widehat{\mathbf{1}}
	\end{array}\right)\,,\; & \overline{\mathbf{6}}=\left(\begin{array}{ccc}
	\widehat{\mathbf{3}}_{-1} & -\frac{1}{\sqrt{2}}\widehat{\mathbf{3}}_{0} & \frac{1}{\sqrt{2}}\widehat{\mathbf{2}}_{-1/2}\\
	-\frac{1}{\sqrt{2}}\widehat{\mathbf{3}}_{0} & \widehat{\mathbf{3}}_{1} & -\frac{1}{\sqrt{2}}\widehat{\mathbf{2}}_{1/2}\\
	\frac{1}{\sqrt{2}}\widehat{\mathbf{2}}_{-1/2} & -\frac{1}{\sqrt{2}}\widehat{\mathbf{2}}_{1/2} & \widehat{\mathbf{1}}
\end{array}\right)\,.\label{eq:6plet}
\end{align}
A mass terms $\textrm{Tr}\left(\overline{\mathbf{6}}'\cdot\mathbf{6}\right)$
then translates into $\textrm{Tr}\left(\overline{\Delta}'\cdot\Delta\right)+\widehat{\mathbf{2}}'_{-1/2}\widehat{\mathbf{2}}{}_{1/2}-\widehat{\mathbf{2}}'_{1/2}\widehat{\mathbf{2}}{}_{-1/2}+\widehat{\mathbf{1}}'\widehat{\mathbf{1}}$
for two $SU(2)_{L}$ triplets $\Delta$ and $\overline{\Delta}'$
which we can write in terms of isospin components as
\begin{align}
	\overline{\Delta}'=\left(\begin{array}{cc}
		\widehat{\mathbf{3}}'_{-1} & -\frac{1}{\sqrt{2}}\widehat{\mathbf{3}}'_{0}\\
		-\frac{1}{\sqrt{2}}\widehat{\mathbf{3}}'_{0} & \widehat{\mathbf{3}}'_{1}
	\end{array}\right)\,,\; & \Delta=\left(\begin{array}{cc}
	\widehat{\mathbf{3}}_{1} & \frac{1}{\sqrt{2}}\widehat{\mathbf{3}}_{0}\\
	\frac{1}{\sqrt{2}}\widehat{\mathbf{3}}_{0} & \widehat{\mathbf{3}}_{-1}
\end{array}\right)\,.
\end{align}

Note that conventions in the literature vary regarding the signs in
front of some of the triplet components $\widehat{\mathbf{3}}_{i}$,
since these might change with a rephasing of fields components. However,
the $\frac{1}{\sqrt{2}}$ factors cannot be absorbed, so the expression
in equation (\ref{eq:6plet}) for the sextet differs in a material
way from the one used in \cite{Foot1993,Tully2001,Okada2016}, for
example, agreeing instead with \cite{Pires2014}.\footnote{Without these $\frac{1}{\sqrt{2}}$ factors, it is easy to check that
	a mass term $\textrm{Tr}\left(\mathbf{6}^{\dagger}\mathbf{6}\right)$
	will not correspond to the sum of the norm-squared of all six components.}

Finally, we consider what happens to gauge bosons $W_{L,i}$ ($i=1,\cdots8$)
which are in the adjoint representation ($\mathbf{8}$) of $SU(3)_{L}$.
The representation $\left(\mathbf{8},0\right)$ of $SU(3)_{L}\times U(1)_{X}$
breaks into one $SU(2)_{L}$ triplet $\widehat{\mathbf{3}}$, one
singlet $\widehat{\mathbf{1}}$ and two doublets $\widehat{\mathbf{2}}$
and $\widehat{\mathbf{2}}'$ with opposite hypercharges; for definiteness
let us consider $\widehat{\mathbf{2}}$ to be the one with $y=\frac{\sqrt{3}}{2}\beta$
--- see equation (\ref{eq:octet_decomposition}). Contractions with
(anti)triplets are done in the standard way ($\mathbf{8}_{ij}\overline{\mathbf{3}}{}_{i}\mathbf{3}_{j}$),
resulting in the following identification of the octet components:
\begin{align}
	\mathbf{8} & =\left(\begin{array}{ccc}
		\frac{1}{\sqrt{6}}\widehat{\mathbf{1}}-\frac{1}{\sqrt{2}}\widehat{\mathbf{3}}_{0} & \widehat{\mathbf{3}}_{1} & -\widehat{\mathbf{2}}_{1/2}\\
		-\widehat{\mathbf{3}}_{-1} & \frac{1}{\sqrt{6}}\widehat{\mathbf{1}}+\frac{1}{\sqrt{2}}\widehat{\mathbf{3}}_{0} & -\widehat{\mathbf{2}}_{-1/2}\\
		-\widehat{\mathbf{2}}'_{-1/2} & \widehat{\mathbf{2}}'_{1/2} & -\sqrt{\frac{2}{3}}\widehat{\mathbf{1}}
	\end{array}\right)\label{eq:octet-1}
\end{align}
In the case of gauge bosons, we are dealing with a \textit{real} field
transforming as $\mathbf{8}$ hence the $SU(2)_{L}$ $\widehat{\mathbf{2}}$
and $\widehat{\mathbf{2}}'$ doublets are not independent. Indeed,
one can alternatively write $\mathbf{8}=\frac{1}{\sqrt{2}}W^{a}\lambda^{a}$,
where $\lambda^{1,\cdots,8}$ are the Gell-Mann matrices:

\begin{align}
	\mathbf{W}_{L} & =\frac{1}{\sqrt{2}}\left(\begin{array}{ccc}
		W_{L}^{3}+\frac{1}{\sqrt{3}}W_{L}^{8} & W_{L}^{1}-iW_{L}^{2} & W_{L}^{4}-iW_{L}^{5}\\
		W_{L}^{1}+iW_{L}^{2} & -W_{L}^{3}+\frac{1}{\sqrt{3}}W_{L}^{8} & W_{L}^{6}-iW_{L}^{7}\\
		W_{L}^{4}+iW_{L}^{5} & W_{L}^{6}+iW_{L}^{7} & -\frac{2}{\sqrt{3}}W_{L}^{8}
	\end{array}\right)\,.\label{eq:octet-2}
\end{align}
Equating the expressions in equations \eqref{eq:octet-1} and \eqref{eq:octet-2}, we get the identification
\begin{align}
	\left(\begin{array}{c}
		\widehat{\mathbf{3}}_{1}\\
		\widehat{\mathbf{3}}_{0}\\
		\widehat{\mathbf{3}}_{-1}
	\end{array}\right)=\left(\begin{array}{c}
	\widehat{\mathbf{3}}_{-1}^{*}\\
	\widehat{\mathbf{3}}_{0}^{*}\\
	\widehat{\mathbf{3}}_{1}^{*}
\end{array}\right) & =\left(\begin{array}{c}
\left(W_{L}^{1}-iW_{L}^{2}\right)/\sqrt{2}\\
-W_{L}^{3}\\
\left(W_{L}^{1}+iW_{L}^{2}\right)/\sqrt{2}
\end{array}\right)\,,\\
\left(\begin{array}{c}
	\widehat{\mathbf{2}}_{1/2}\\
	\widehat{\mathbf{2}}_{-1/2}
\end{array}\right)=\left(\begin{array}{c}
\widehat{\mathbf{2}}_{-1/2}^{\prime*}\\
-\widehat{\mathbf{2}}_{1/2}^{\prime*}
\end{array}\right) & =\left(\begin{array}{c}
\left(-W_{L}^{4}+iW_{L}^{5}\right)/\sqrt{2}\\
\left(-W_{L}^{6}+iW_{L}^{7}\right)/\sqrt{2}
\end{array}\right)\,,\\
\widehat{\mathbf{1}} & =W_{L}^{8}\,.
\end{align}
It is then obvious that the Standard Model $SU(2)_{L}$ gauge bosons
correspond to the triplet $\widehat{\mathbf{3}}$ (i.e., $W_{L}^{1,2,3}$)
while the singlet $\widehat{\mathbf{1}}$ (i.e., $W_{L}^{8}$) mixes
with the $U\left(1\right)_{X}$ gauge boson $W_{X}$ to form the $U(1)_{Y}$
gauge boson $B$:
\begin{align}
	B & =\frac{1}{\sqrt{g_{L}^{2}+g_{X}^{2}\beta^{2}}}\left(g_{X}\beta W_{L}^{8}+g_{L}W_{X}\right)\,.
\end{align}
In this expression, $g_{L}$ and $g_{X}$ stand for the gauge coupling
constants of $SU(3)_{L}$ and $U\left(1\right)_{X}$, which are related
to $g_{Y}$ through the relation\footnote{The relation changes if we choose instead to normalize the $X$ and
	$Y$ charges in a different way \cite{Boucenna2015}. }
\begin{align}\label{eq:gY}
	g_{Y}^{-2} & =\beta^{2}g_{L}^{-2}+g_{X}^{-2}\,.
\end{align}

Finally, note that the charge of the various components of $\mathbf{W}_{L}$
depend only on $\beta$: 
\begin{align}
	Q\left(\mathbf{W}_{L}\right)_{\beta} & =\left(\begin{array}{ccc}
		0 & + & 0\\
		- & 0 & -\\
		0 & + & 0
	\end{array}\right)_{\!\!\!-\frac{1}{\sqrt{3}}},\,\left(\begin{array}{ccc}
	0 & + & +\\
	- & 0 & 0\\
	- & 0 & 0
\end{array}\right)_{\!\!\!\frac{1}{\sqrt{3}}},\,\left(\begin{array}{ccc}
0 & + & -\\
- & 0 & --\\
+ & ++ & 0
\end{array}\right)_{\!\!\!-\sqrt{3}},\,\left(\begin{array}{ccc}
0 & + & ++\\
- & 0 & +\\
-- & - & 0
\end{array}\right)_{\!\!\!\sqrt{3}}\,.
\end{align}

\end{document}